\documentclass{aa}
\usepackage{graphicx}
\usepackage{lscape}
\usepackage{longtable}
\usepackage{txfonts}
\usepackage{natbib}

\begin{document}

   \title{Fe XIII coronal line emission in cool M dwarfs\thanks{Based on observations
collected at the European Southern Observatory, Paranal, Chile, 68.D-0166A.}}

   \titlerunning{Fe XIII in cool M stars}

   \author{B. Fuhrmeister,
          J. H. M. M. Schmitt \and R. Wichmann
          }

   \offprints{B. Fuhrmeister}

   \institute{Hamburger Sternwarte, University of Hamburg,
              Gojenbergsweg 112, D-21029 Hamburg\\
              \email{bfuhrmeister@hs.uni-hamburg.de}
             }


   \abstract{We report on a search for the Fe\,{\sc xiii} forbidden coronal line at 33881.1 \AA \ 
             in a
             sample of 15 M-type dwarf stars covering the whole spectral class as well as different 
             levels of activity. A clear detection was achieved
             for LHS 2076 during a major flare and for CN~Leo, where the line had
             been discovered before. For some other stars the situation is not quite 
             clear. For CN~Leo we investigated the timing behaviour of the Fe\,{\sc xiii}
             line and report a high level of variability on a timescale of hours
             which we ascribe to microflare heating.

   }

   \maketitle
%

\section{Introduction}

The solar corona can be studied at extremely high spatial and spectral
resolution over almost the entire range of the electromagnetic spectrum.
With typical coronal plasma temperatures of a few million degrees 
the bulk of the Sun's coronal energy 
losses is emitted in the X-ray range,
the energy losses at shorter and longer wavelengths being considerably
smaller. Therefore observations of the solar corona in the optical
are quite difficult because of the faintness of 
the corona compared to the photosphere in this wavelength range.  
This problem can be overcome at least for observations
above the solar limb if the photospheric light is blocked, e.g. by the Moon 
during an eclipse or -- artificially -- by a coronograph.  Optical observations 
of the solar corona carried out in that fashion were the first to 
reveal the true nature of the corona as a hot plasma with temperatures much 
higher than the underlying chromosphere or photosphere.  However, the solar 
corona could only be observed off the limb in the optical, and only 
through satellite-based imaging observations of the corona at X-ray and EUV 
wavelengths the corona at large could be observed and studied.

The existence of coronae similar to that of the Sun around essentially
all late-type main sequence stars with outer convection zones has been
established by extensive X-ray studies (e.\,g. \citet{Schmitt1}).  
As is the case for the Sun, the bulk of the energy losses for stellar 
coronae also occurs in the X-ray range, and consequently this
spectral band is the most natural one for the study of stellar coronae.  However, 
coronal emission from stars can also be observed in the radio range (e.\,g. \citet{Berger} 
or \citet{Guedel}) and
in the optical (\citet{nature}).   Coronal observations at those latter wavebands are 
extremely difficult since the 
coronal emissions are much fainter, and especially in the optical, the weak coronal
emission has to be detected above the usually much brighter optical photospheric
emission.  The most promising candidates to search for optical coronal emission 
are clearly late-type M dwarf stars, which can be as X-ray bright as or even 
X-ray brighter than the Sun, but whose photospheric emission is rather faint in 
particular at near UV wavelengths.
The detection of coronal emission in the optical was recently accomplished for 
the active M-star CN~Leo, where \citet{nature} were able to detect
the Fe\,{\sc xiii} forbidden coronal line at 3388.1 \AA. This successful detection
of coronal emission in one star raises the question to what extent such detections are 
possible for other cool stars as well or whether CN~Leo is a unique and singular case.  

In this paper we will discuss the problems of detecting the 3388.1 \AA\, forbidden coronal line in a small 
sample of late-type stars and present an analysis of
the temporal variability of this line in the "proto-type" CN~Leo.  
Our paper is structured as follows: 
In section 2 we describe the VLT data used for our 
analyses and the sample of investigated stars. 
In section 3 an overview over the
spectral range under investigation is given, while in section 4 we deal with 
the rotational velocity of the analyzed stars. In section 5 the results of our search
for the Fe\,{\sc xiii} line for individual stars are presented and
the timing behavior of the line for CN~Leo is
discussed in section 6. Section 7 deals with the X-ray to Fe\,{\sc xiii} line ratio for
CN~Leo and LHS~2076. In section 8 we describe the results of our search for other forbidden
coronal lines in CN~Leo.


\section{Observations and data analysis}

The observations reported in this paper were obtained in visitor mode with 
ESO's Kueyen telescope at Paranal equipped with the Ultraviolet-Visual Echelle 
Spectrograph (UVES) from March, 13th to 16th in 2002, with the exception of 
UV~Ceti, Prox~Cen, LHS 292 and one of the CN~Leo spectra, 
which were observed in service mode during the winter season 2000/2001.  
For the March 2002 run the instrument 
was operated in a dichroic mode, yielding
33 echelle orders in the blue arm (spectral coverage from 3030 to 3880 \AA) and 39 orders
in the red arm (spectral coverage from 4580 to 6680 \AA). For the runs in the
winter 2000/2001 a monochroic setup was used, providing us only with the blue part of 
the spectrum. In the dichroic setup the red part of the
spectrum is recorded on two separate CCDs; therefore there is a spectral gap from 
$\sim$ 5640 to 5740 \AA, resulting from a spatial gap between the two CCDs. As a consequence
of the various instrumental constraints we cannot observe the lines from H$_{3}$ up to H$_{8}$ of the Balmer series,
nor do we cover the Ca\,{\sc ii} H and K lines.  
The typical resolution of our spectra is $\sim 45000$, typical exposures times were 20 minutes 
except for the brighter among our sample stars. Normally for each star three exposures were
taken in sequence in order to facilitate the recognition of cosmics, again with the exception of the
four spectra of CN~Leo, UV~Ceti, Prox~Cen and LHS 292 taken during the winter 
season 2000/2001.

A list of the observed stars and the available spectra for each star is 
provided in Table \ref{starlist}, giving the observation dates
and total exposure times. Moreover we provide some basic parameters 
of the sample stars,  including the highest detected Balmer line in our data 
as an estimator of activity.  Note that during the March 2002 run CN~Leo was 
observed nightly to carry out an investigation of its chromospheric and coronal variability.

All data were reduced using IRAF, including flat-fielding, order definition and scattered light
subtraction. The wavelength calibration was carried out using Thorium-Argon spectra with
an accuracy of $\sim 0.03$ \AA\,in the blue arm and $\sim 0.05 $\AA\,in the red arm.
In addition there are photometric data from the UVES exposure meter 
taken for engineering purposes and therefore not flux calibrated. 
Still, these data were useful to assess 
whether the star was observed during quiescence or during a major flare. 

\begin{table*}[!ht]
\caption{\label{starlist}Basic observations parameters of the observed stars.}
\begin{tabular}{lllllll}
\hline
name & other & spectral  & log $\mathrm{L_{X}} ^{1}$ & observations &Ti\,{\sc ii} lines&  highest\\
     &  name & type &                     &         &      & Balmer line\\
\hline
LHS 1827& GJ 229A&M1 &27.13$^{a}$ & 2002-03-15 4 spectra 1200s& absorption& -\\
LHS 428 &&M3e &28.75$^{a}$ & 2002-03-15 2 spectra 1200s& absorption& H$_{18}$\\
LHS 6158 &  &M3.5 &28.76$^{c}$ & 2002-03-15 2 spectra 2400s& weak absorption&H$_{18}$\\
LHS 5167 & AD~Leo & M3.5 &28.92$^{a}$ & 2002-03-13 3 spectra 1800s& emission& H$_{18}$\\
         &        &      & &2002-03-16 2 spectra 1200s&    &     \\
HD 196982 & AT~Mic & M4.5 &29.55$^{c}$ & 2002-03-16 2 spectra 2400s& emission& H$_{20}$\\ 
LHS 1943 & YZ~CMi & M4.5e&28.67$^{b}$ & 2002-03-13 3 spectra 3600s& emission& H$_{24}$\\
LHS 2664 & FN~Vir & M4.5 &27.92$^{b}$ & 2002-03-13 3 spectra 3600s& weak emission&H$_{18}$\\
LHS 324 & GL~Vir & M5 &27.65$^{a}$ & 2002-03-13 3 spectra 3600s& weak emission& H$_{19}$\\
        &        &    & & 2002-03-16 2 spectra 2400s&      &   \\
LHS 36 & CN~Leo & M5.5 &27.78$^{a}$ & 2002-03-13 6 spectra 7200s& emission& H$_{24}$\\
       &        &      & & 2002-03-14 4 spectra 4800s&      &  \\
       &        &      & & 2002-03-15 6 spectra 7200s&      & \\
       &        &      & & 2002-03-16 6 spectra 7200s&        & \\
       &        &      & & 2001-01-06 1 spectrum 3120s&      &  \\
LHS 2076 &EI~Cnc & M5.5 &27.60$^{a}$ & 2002-03-15 4 spectra 4800s& - & H$_{18}$\\
         & &      & & 2002-03-16 1 spectrum 1200s& & \\
LHS 49 & Prox~Cen & M5.5 &27.26$^{a}$ & 2001-02-02 1 spectrum 3120s& weak emission& H$_{18}$ \\ 
LHS 10 & UV~Cet & M5.5 &27.31$^{a}$ &2000-12-17 1 spectrum 3120s& emission & H$_{18}$\\
LHS 248 & DX~Cnc &  M6 &26.60$^{a}$ & 2002-03-16 3 spectra 3600s & - &H$_{11}$\\
LHS 2034 &AZ~Cnc & M6 &28.89$^{c}$ & 2002-03-14 6 spectra 6000s& emission& H$_{17}$\\
         & &    & & 2002-03-16 2 spectra 2400s&       &  \\
LHS 292 &  & M6.5&   &2001-02-02 1 spectrum 3120s& - &H$_{15}$\\

\hline
\end{tabular}
\\
$^{1}$ in erg\, s$^{-1}$\\
$^{a}$ \citet{low-massx}\\
$^{b}$ \citet{Delfosse}\\
$^{c}$ \citet{Huensch}\\
$^{d}$ \citet{lhs2065} X-ray luminosity is variable
\end{table*}

The spectral line fits were carried out with the CORA fit program \citep{cora}. 
This software tool was originally developed for analyzing high resolution X-ray spectra, 
but the fit algorithms employed
by CORA are also well suited for the modeling of well-defined chromospheric and 
coronal emission lines. 
The program  provides an accurate error analysis. 
All fits were carried out using Gaussian line profiles after shifting
the wavelength to the stars' rest frame.  


\section{Detection of the Fe\,{\sc xiii} line}

The prime target of our observing program was the spectral region around 3388 \AA.
The specific transition under consideration is  $3s^{2}3p^{2}$\ $^{3}P_{2} - ^{1}\!D_{2}$. 
\citet{Flower} 
give a detailed discussion of the atomic physics
of Fe\,{\sc xiii} and provide level population calculations for a range of
temperatures and densities, considering direct excitation due to collisions with 
electrons and protons and indirect excitation of allowed transitions via collisions 
with electrons followed by radiative decay and
find rather low dependence on temperature.

In order to give an overview of the nature and quality of
our spectra, we show (in Fig. \ref{range}) the spectral range between 3370 \AA\,- 3390 \AA\, 
for most of our program stars.  LHS 292 is not shown since its spectrum is very similar to DX~Cnc,
and LHS 428 and LHS 6158 are not shown since they are both double stars with very complicated
spectra that could not be disentangled and were therefore excluded from the analysis.
 The spectra shown in Fig. \ref{range} are sorted by decreasing spectral 
type.  
One notes a distinct sequence from ``early'' M dwarfs to ``late'' M dwarfs.  In the
``early'' M dwarfs like Gl 229A and AD Leo one still recognizes a clear photospheric spectrum in the
range  3370 \AA\,- 3390 \AA\, with some additional chromospheric emission 
lines,  while in the later type M 
dwarfs like CN Leo and the (flaring) LHS~2034 an almost pure emission line 
spectrum appears; these latter spectra are obviously dominated by the stellar chromospheres 
rather than their photospheres.  

The Fe\,{\sc xiii} line at 3388.1 \AA\, detected in CN~Leo by \citet{nature} is 
-- unfortunately -- blended with a Ti\,{\sc ii} chromospheric line at 3387.846 \AA.
This particular Ti\,{\sc ii} line and other emission lines from the same 
multiplet are present in all recorded spectra besides 
the non-flare spectra of LHS 2034 and LHS 2076 and the spectrum of 
DX~Cnc.  
 In GL 229A Ti\,{\sc ii} appears to be
present in absorption.
 
The stars earlier than M4.5 have a
rather high pseudo-continuum in the considered wavelength range which is
clearly dominated by overlapping absorption lines. This continuum strongly decreases for the 
spectral types 
M3.5 - M4.5 stars by a factor of ten, accompanied by the vanishing of the strong
absorption features in this regime. Consequently, a detection of the Fe\,{\sc xiii} line should be 
much easier 
for spectral types later than M4.5. Therefore the most promising search for coronal line 
emission can be made for stars in the
spectral range M4.5 - M6 since for later stars even the normally strong chromospheric
Ti\,{\sc ii} emission lines cannot be detected any more.

\begin{figure*}
\begin{center}
\includegraphics[height=23cm,bbllx=50,bblly=175,bburx=470,bbury=775]{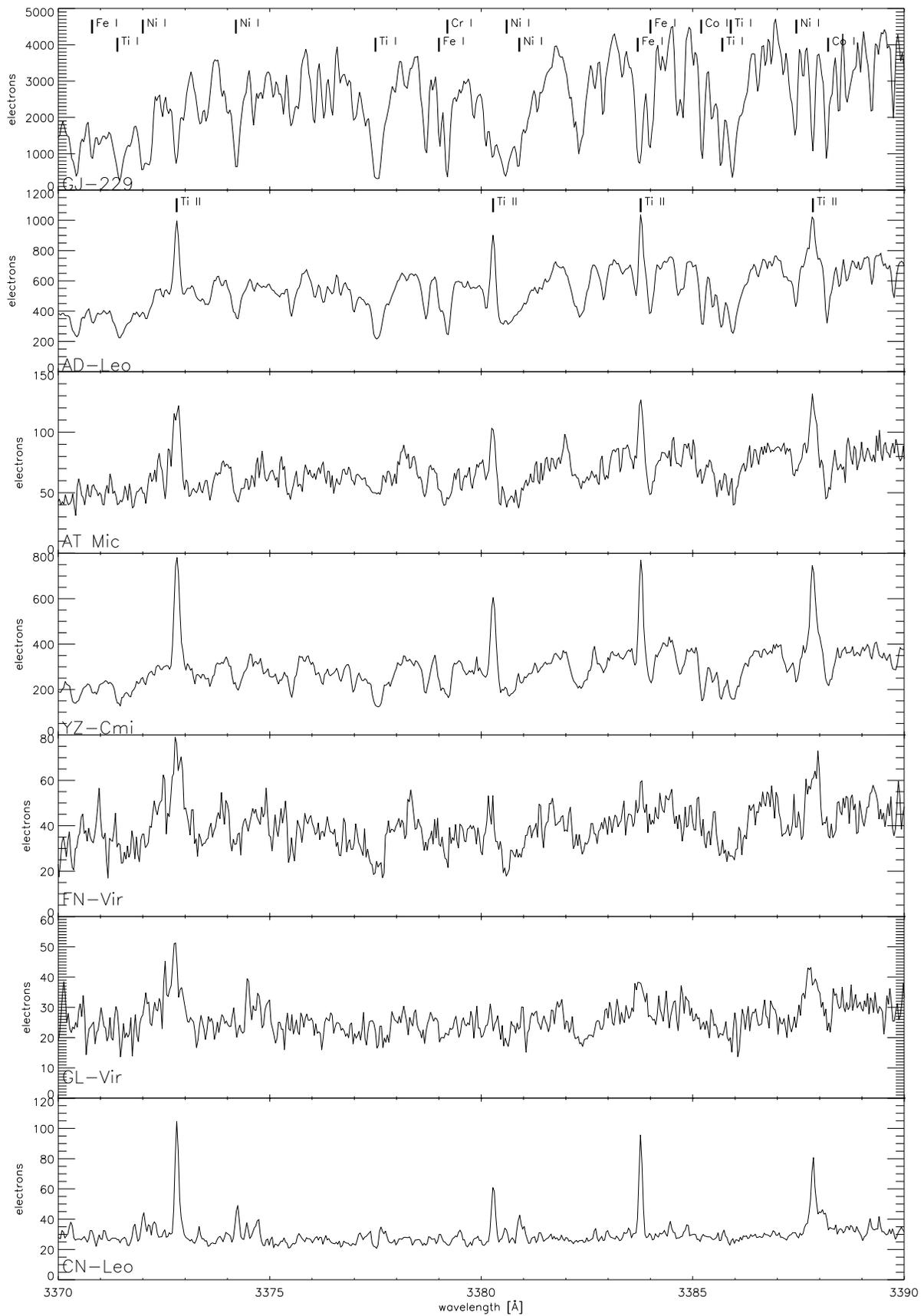}
\caption{\label{range}The considered wavelength range of selected program stars. Some strong 
absorption
lines are identified in the top spectrum. The Ti\,{\sc ii} emission lines are marked in the spectrum
of Ad~Leo.}
\end{center}
\end{figure*}

\setcounter{figure}{0}
\begin{figure*}
\begin{center}
\includegraphics[height=23cm,bbllx=50,bblly=175,bburx=470,bbury=775]{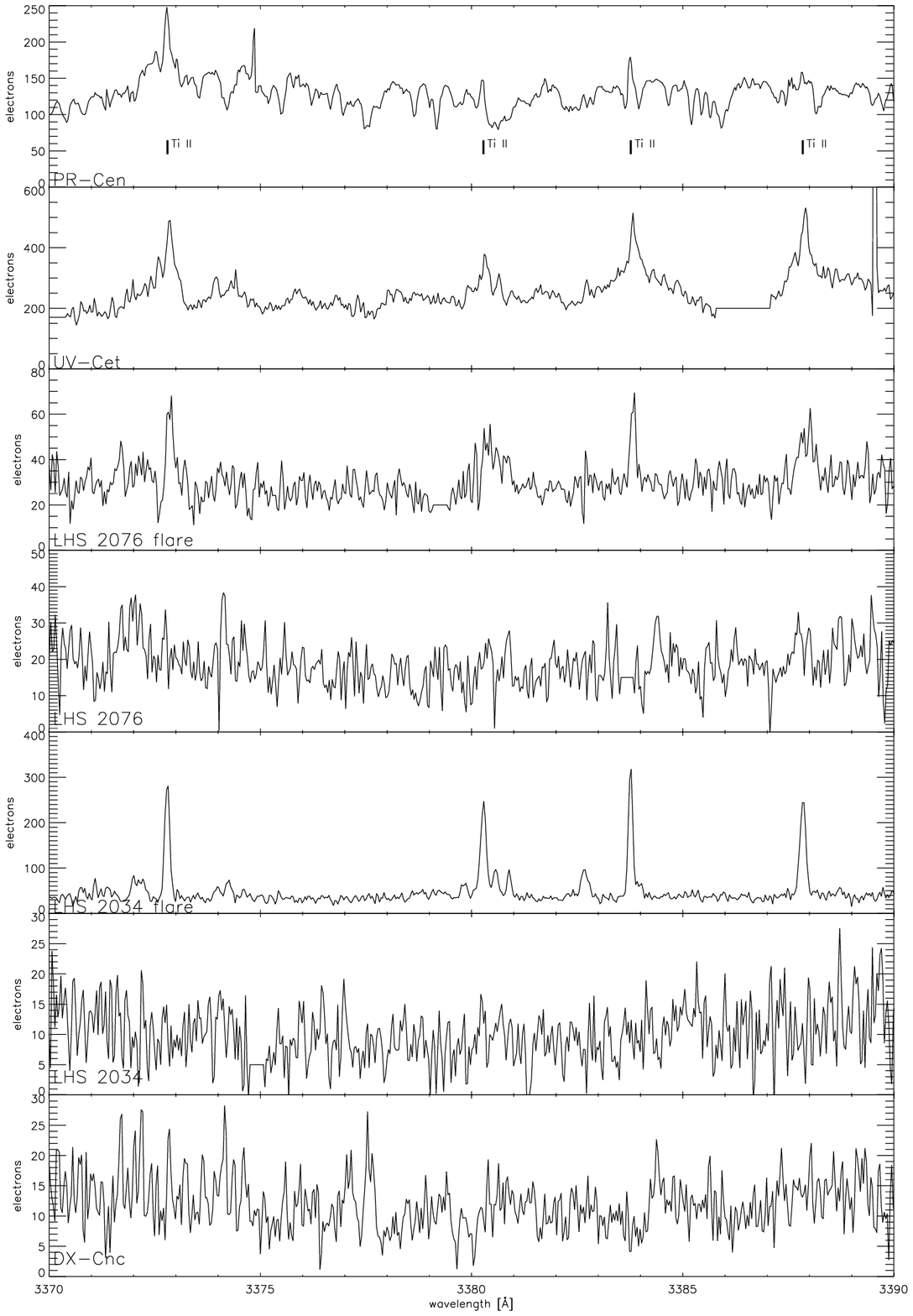}
\caption{(continued) For LHS 2034 and LHS 2076 the spectra are shown during flare and during
quiescent state. Note that these four spectra and the spectrum of UV~Cet are not averaged spectra and
therefore some remaining cosmics were replaced manually by a straight line. }
\end{center}
\end{figure*}


\section{Determination of rotational velocities}\label{rot}

A critical issue for the successful separation of the chromospheric Ti~II emission line
at 3387.846 \AA \ and the coronal Fe~XIII emission line at  3388.1 \AA \ are the intrinsic 
line broadening 
profiles of these lines.   The wavelength difference
between those lines ($\approx$ 0.26 \AA ) corresponds to a velocity of 
$\approx$ 23~km/sec, easily within the reach of thermal and rotational
velocities.  Obviously, the thermal
broadening profiles differ substantially because of the different origins of
these lines, and rapid rotation is expected at least for
the more active stars.  Since we clearly need accurate values for $\mathrm{v}\sin(i)$ for disentangling 
the Ti~II and Fe~XIII emission lines,
we decided to measure it for some of our program stars.  In order to
determine  $\mathrm{v}\sin(i)$ we used 18 of the red arm orders that show no
strong emission lines either from the star or from terrestrial airglow.  As a
template we decided to use CN~Leo since CN~Leo is known to rotate very slowly ($\mathrm{v}\sin(i) < 2.9  \mathrm{km s^{-1}}$
\citep{Delfosse}).  The measured spectrum of CN~Leo was spun up with rotational velocities from 3 up to 45 
$ \mathrm{km s^{-1}}$. The best fit value for the rotation velocity $\mathrm{v}\sin(i)$  of the spun up 
template and the 
star under study was determined with a $\chi^{2}$-test for every order used.  The final rotational velocity was 
determined by averaging the values obtained in the different orders. We mostly used only the red part of 
the spectrum 
since in the blue part of the spectrum it is rather difficult to find larger regions
unaffected by emission lines. 
We tried to avoid this problem of the blue spectra by
using only very narrow spectral ranges ($\sim 10$\AA) of every aperture, yet the issues of 
which lines are in
emission, which show emission cores and which are purely in absorption are very different for 
each program star. Therefore the template
has to be very similar to the test star and the used wavelength intervals have to be determined
individually but even then the scatter for the wavelength intervals is  bigger than in
the red arm. Therefore we determined the rotational velocity for only three stars in the blue arm.
For these stars we used Prox Cen as template since its spectrum is very similar to these stars. CN~Leo
could not be used as template because it is much more active than the measured stars.
The rotational velocities determined in this fashion are listed in Table \ref{rotation}.  In general, 
the values determined by us agree with previously determined
ones found in the literature with the exception of GL~Vir, where we
find a substantially larger rotational velocity both in the red and blue arm
than \citet{Delfosse}. A careful visual inspection and comparison of the spectra of
GL~Vir with those obtained for FN~Vir and AD~Leo shows a very high similarity to 
the faster rotating spectrum. 
Moreover we used a template star very similar to GL~Vir in spectral type while
\citeauthor{Delfosse} used a synthetic K0III spectrum.  
Therefore we do favor our (larger) rotational velocity.

For the double star UV~Cet we measure a rotational velocity 
of 25.9 $\pm$ 8.6 $\mathrm{km\,s^{-1}}$ which
agrees with the value found by \citet{Mohanty} for UV~Cet B. Since UV~Cet A
has a similar rotation velocity (G. Basri, private communication) it cannot be
decided from the rotational velocity which component of UV~Cet dominates our spectrum.

\begin{table}
\caption{\label{rotation}Rotational velocities for (some of) our program stars. 
LHS 428 is omitted
because it is a double star  and could not be separated. LHS 6158 even seems to 
be a triple system.
GJ 229A is saturated in the red arm. Prox~Cen was used as template for UV~Cet, FN~Vir and GL~Vir
in the blue arm. All measurements in $\mathrm{km\,s^{-1}}$.}
\begin{tabular}[htbp]{cccc}
\hline
star & $\mathrm{v}\sin(i)$ & $\mathrm{v}\sin(i)$ & literature\\
     & red arm & blue arm &  \\
\hline

AD Leo & 7.6 $\pm$ 1.2&  & 6.2 $\pm$ 0.8$^{1}$\\
AT Mic & 11.7 $\pm$ 0.9& &  \\
YZ CMi & 6.7 $\pm$ 0.6 & & 7.0 $\pm$ 2.0$^{2}$, 6.5$\pm$ 1.1$^{1}$\\
FN Vir & 17.4 $\pm$ 1.4 & 13.1 $\pm$ 3.7 & 16.8 $\pm$ 2.1$^{1}$\\
GL Vir & 15.3 $\pm$ 0.8 & 16.0 $\pm$ 2.0 & 9.2 $\pm$ 1.9$^{1}$\\
UV Cet B &  & & 32.5 $\pm$ 2.0 $^{2}$\\
UV Cet A & -  & 25.9 $\pm$ 8.6 & \\
DX Cnc & 8.9 $\pm$ 0.8& & 11.0 $\pm$ 2.0$^{2}$, 8.1 $\pm$ 1.1$^{1}$\\ 
LHS 2034 &7.9 $\pm$ 2.8& & \\

\hline
\end{tabular}
\\
$^{1}$ \citet{Delfosse}\\
$^{2}$ \citet{Mohanty}
\end{table}


\section{Observations of individual stars}

Since the spectral properties of our sample stars are very different we applied specific 
methods for the individual
stars for a detection or non-detection of the Fe\,{\sc xiii} line. These are 
discussed in detail below. 

\subsection{CN Leo}

We report a clear detection of the Fe\,{\sc xiii} line for CN~Leo, which is 
also found to be variable on a timescale of hours, as will be discussed 
in section \ref{timeseries}. Thus the Fe~XIII line detection of \citet{nature} is 
fully confirmed.

\subsection{LHS 2076}\label{corafitting}

\begin{figure*}[hb!t]
\begin{center}
\includegraphics[width=8cm,height=5cm,clip=]{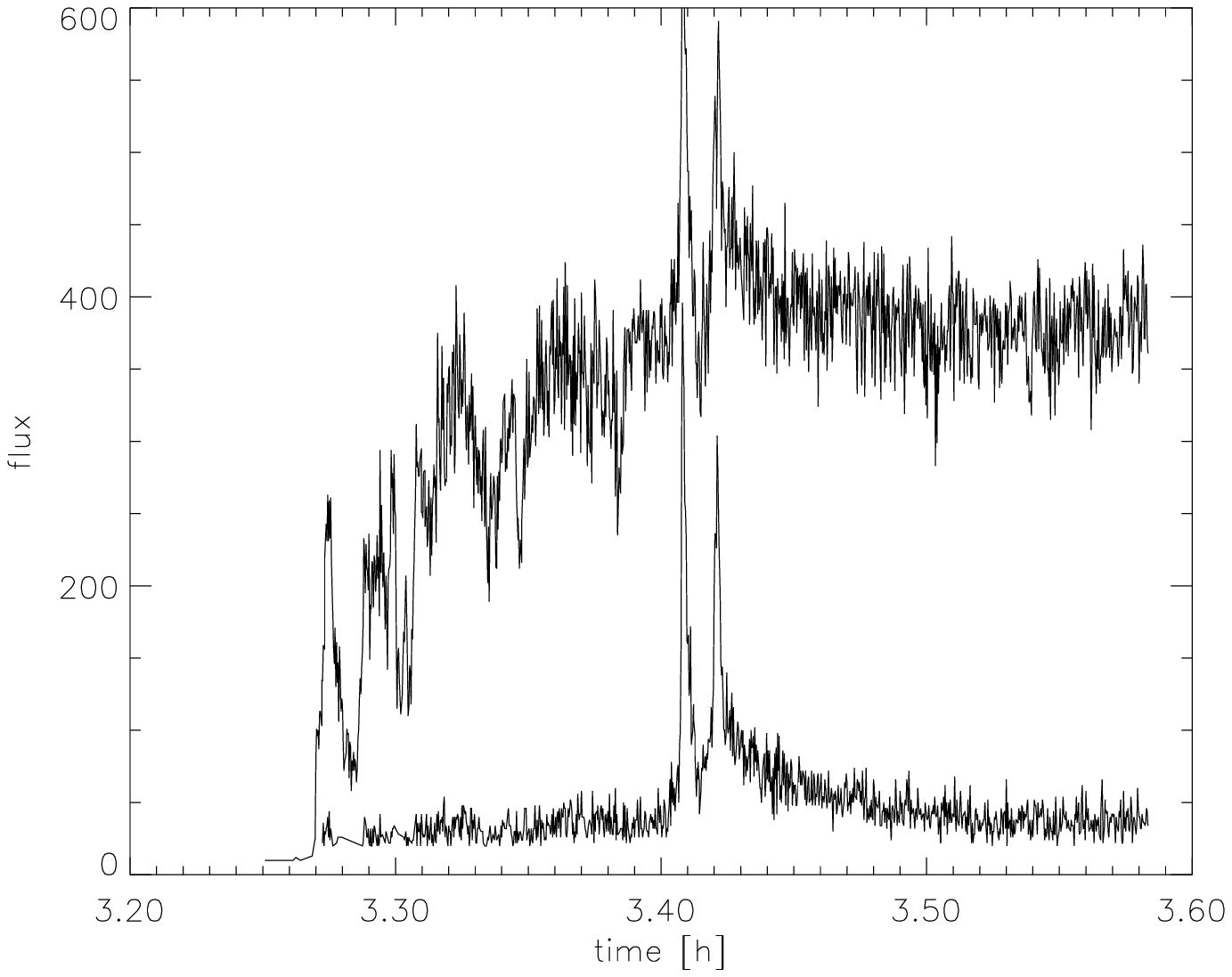}
\includegraphics[width=8cm,height=5cm,clip=]{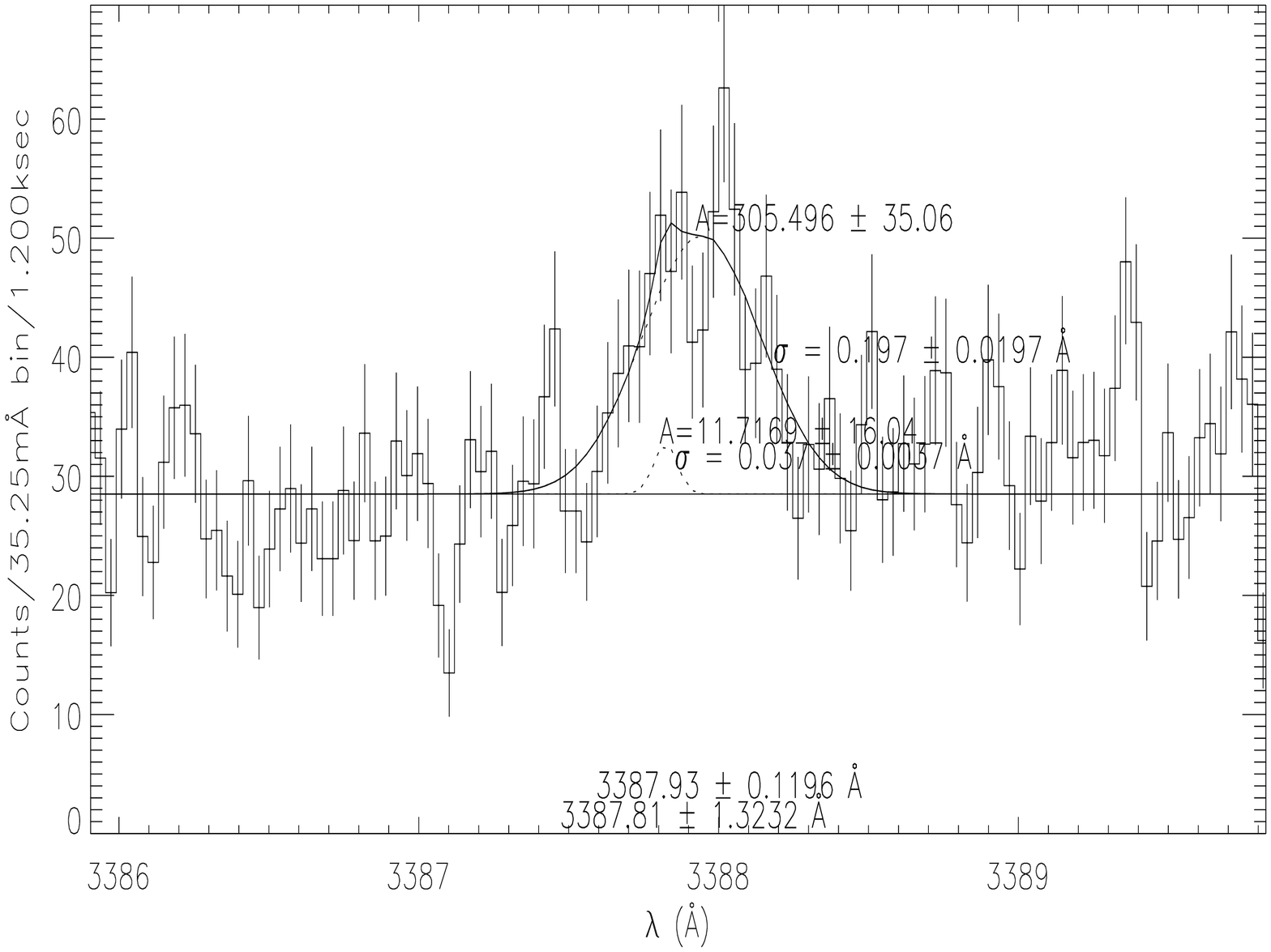}
\caption{\label{photm}To the left the 
light curve of the flux of LHS 2076 on March, 15th, corresponding to a
single spectrum. On the x-axis the universal time is given. The flux is in arbitrary units since
the photometer is used for engineering purposes. The upper light curve corresponds to
the red arm of the spectra, while the lower light curve corresponds to the blue arm and 
is scaled relative to the red flux for convenience.
Two short duration flares can be seen 
in both the red and the blue arm of the photometer following each other rapidly.
To the right the spectrum of LHS 2076 during the flare around 3388 \AA\, showing an
broad emission feature at 3387.92 \AA.}
\end{center}
\end{figure*}

A clear detection of the Fe\,{\sc xiii} 3388.1 \AA\,\ emission line has been obtained during
a flare on  LHS~2076.  LHS~2076 is a double star known for its flare activity and 
separated by $3^{\prime\prime}.4$ with both components being late-type \citep{Pettersen}.  
Our spectrum refers to one system component with some minor contribution from the second star. 
During the last of
the four exposures on March, 15th a double-peaked flare can be recognized in the photometer 
light curve as shown in Fig. \ref{photm}. 

Outside the
flare interval no significant Ti\,{\sc ii} emission is detectable
in the spectra. The
emission line properties of the flare spectrum are listed
in Table \ref{lhs2076}. 
The emission feature at 3387 \AA\ \ is quite broad and was fitted with a 
double line component with amplitude, central wavelength 
and FWHM for both components as free parameters. 
The other Ti\,{\sc ii} lines were fitted as single lines with all three 
parameters ($\mathrm{\lambda_{cen},} 
\mathrm{\,A\,\, and\,\, }  \sigma$) free.  Inspecting the properties of these three Ti\,{\sc ii} lines 
(cf., Table \ref{lhs2076}) one finds that the fitted half widths of all three lines 
agree very well with each other as expected for lines from the same multiplet.  
The half width of the narrow component of the 3387 \AA\, line also agrees well 
with the half widths of the three Ti\,{\sc ii} lines while the half width of the broad 
component is more than three times larger. We
therefore conclude that the narrow component of the broad emission feature at
3387 \AA\ must be the fourth Ti\,{\sc ii} line in the
multiplet, while the broad component has to be identified with the
Fe\,{\sc xiii} line.
Further evidence for this interpretation can be found from the lines' amplitudes.  
Atomic data available 
through the NIST atomic spectra database \footnote{available via \\ 
http://physics.nist.gov/cgi$-$bin/AtData/main\_asd} for
these lines predict equal relative intensities for the two lines at 3372.80 
and \mbox{3383.77 \AA}\, and
equal relative intensities for the two lines at 3380.28 and 3387.84 \AA\, lowered 
by a factor of four compared to the former two lines. 
According to Table \ref{lhs2076} the amplitudes of the 3372.80 and
3383.77 \AA\, line agree to within the errors and the two other lines are weaker.
Therefore the broad component cannot be attributed to Ti\,{\sc ii} since its
amplitude is more than twice as large as that of the 3372.80 or 3383.77 \AA\, lines. 
On the other hand, both amplitudes and half widths fit perfectly if the broad
emission component is identified the Fe\,{\sc xiii} line; also, the fitted
half widths agree very well with the half width determined for the Fe\,{\sc xiii} 
line in the CN~Leo time series (see section \ref{timeseries}). The central 
wavelength of the absolute line position is a bit on the 
 blue side but well within the 2-$\sigma$ error, while the absolute positions 
of the three single fit Ti\,{\sc ii} lines 
seem to be slightly redshifted (but only to within 1-$\sigma$ error each). 
The 3387.84 \AA\, Ti\,{\sc ii} line unfortunately has
a large error in its central wavelength.  Given this situation, we cannot draw
any meaningful conclusions about possible wavelength shifts and remark that 
during a flare line shifts frequently occur due to mass motions.  

\begin{table*}
\caption{\label{lhs2076}Properties of the Ti\,{\sc ii} and the Fe\,{\sc xiii} line during the 
flare on LHS 2076. Given are the amplitude (in electrons), the central wavelength 
and half width $\sigma$
of each line for the best fit. For the 3380.28 \AA\, line no error estimation for the half width
was possible due to the low signal to noise ratio.}
\begin{tabular}[htbp]{cccccc}
\hline
 &Ti\,{\sc ii} (3372.80 \AA) & Ti\,{\sc ii} (3380.28 \AA) & Ti\,{\sc ii} (3383.77 \AA) & Ti\,{\sc ii} (3387.84 \AA) &
Fe\,{\sc xiii} (3388.1 \AA)\\
\hline
amplitude&134.3$\pm$19.2&67.3$\pm$17.3&115.1$\pm$17.3&11.7$\pm$16.0&305.5$\pm$35.1\\
central $\lambda$& 3372.85$\pm$0.05&3380.33$\pm$0.15&3383.81$\pm$0.04&3387.81$\pm$1.3&3387.93$\pm$0.1\\
$\sigma$&0.06$\pm$0.01&0.06&0.05$\pm$0.01&0.04$\pm$0.01&0.20$\pm$0.02\\
\hline
\end{tabular}
\end{table*}

\subsection{GL Vir}

The continuum for GL~Vir is low as for CN~Leo, but unlike CN~Leo the Ti\,{\sc ii} at 3380 \AA\, line 
is barely detectable.
Nevertheless there is a broad emission line at \mbox{3387.81$\pm$0.14 \AA}\, with a half width
of 0.15 \AA\, and an amplitude of 141.7$\pm$28.9 counts. If this emission
is attributed to the Ti\,{\sc ii} line, the measured amplitude would contradict 
the atomic data, which predicts the same intensity as for the
3380 \AA\, line (which is not seen at all, cf, Fig. \ref{range}). If, on the other hand,  this 
feature is identified with
the Fe\,{\sc xiii} line, it is clearly blue shifted. Since the star is in a quiescent state 
(as can be seen from the light curve, which is not shown here)
it is unlikely that a blue shift is caused by mass motions. However, the blue shift could 
alternatively be caused by absorption at the red wing of the Fe\,{\sc xiii} line due 
to a Co I line at 3388.17 \AA\, which is also seen in other spectra 
(see Fig. \ref{range}). 
More evidence for an interpretation involving the Fe\,{\sc xiii} line can be gained 
from the amplitude of the 3383.71 \AA\, line of 65 $\pm$ 15 counts. 
The 3387 \AA\, Ti II line should have a smaller amplitude according to the atomic data 
and the measured spectra of CN~Leo.
On the other hand, the half widths of the 3383.71 \AA\, line (0.13$\pm 0.02$ \AA) 
and of the 3387.81 \AA\,  line (0.15 $\pm 0.01$) are consistent with each other. 
The Ti II line at 3372.73 \AA\, is unfortunately too deformed to allow any conclusions 
on its half width or amplitude. Therefore the question 
arises whether the half widths of 
the two lines at 3383.71 \AA\, and  3387.81 \AA\, can be caused by rotation.
The rotation velocity $\mathrm{v}\sin(i)$ for GL~Vir was measured by \citet{Delfosse} as 
9.2 $\pm$ 1.9 $\mathrm{km\, s^{-1}}$, while we determined
15.3 $\pm$ 0.8 $\mathrm{km\, s^{-1}}$ (see section \ref{rot}). Thus, if our
rotational velocity is correct, the half widths $\sigma$ can be explained by rotational
broadening, however, if $\mathrm{v} \sin(i)=$9.2 $\mathrm{km\, s^{-1}}$, 
the measured half width is a somewhat too large to be explained by
the rotation. So the data of GL~Vir are not unambiguous.

\subsection{LHS 2034}

The flare star LHS 2034 was observed during a longer flare on March, 14th, lasting 
over half an hour as can be seen from the light curve of the photometer (not shown here).
While only the very strongest chromospheric emission lines can be seen in 
quiescence, a multitude of emission lines becomes visible in the spectrum taken 
during flare maximum. Indeed, the flare spectrum of LHS~2034 is very
similar to the spectrum of CN~Leo with a rather flat continuum without strong absorption lines
and dominated by strong chromospheric emission lines.  Surprisingly, however, we did not find 
any evidence for Fe\,{\sc xiii} emission in this spectrum. The same analysis
procedure as applied for LHS~2076 was used for the flare spectrum of
LHS~2034. All four Ti II lines are found to have about the same FWHM as expected but also the 
same amplitude contrary to expectation from atomic data and what was found for CN~Leo and LHS~2076. 
This fact remains unexplained.  In particular, we cannot attribute
-- as in LHS~2076 -- the excess emission to Fe~XIII, since this would contradict the
measured FWHM.  Obviously,  efficient chromospheric heating of LHS~2034 does occur
during the flare, but the corona is either not hot enough or too hot for producing a 
detectable Fe\,{\sc xiii} line flux.

\subsection{DX Cnc and LHS 292}
\begin{figure}
\begin{center}

\includegraphics[width=8cm,height=5cm,clip=0]{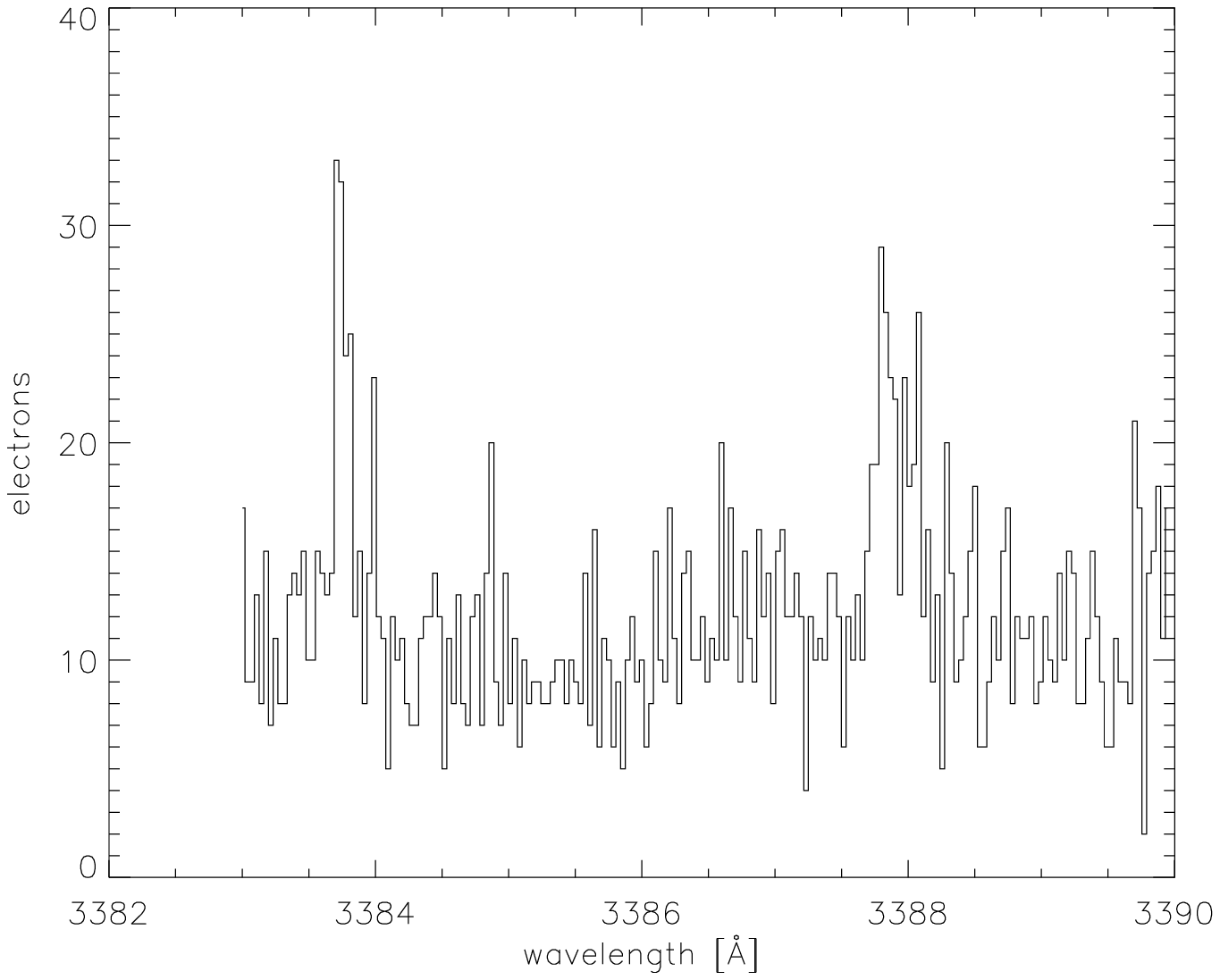}
\caption{\label{sn}Spectrum of CN~Leo with a signal to noise ratio 
artificially reduced to that of DX~Cnc. But despite the low signal to noise ratio
clearly
the Ti\,{\sc ii} and Fe\,{\sc xiii} lines can be recognized.}
\end{center}
\end{figure}

These two stars and LHS~2034 are the latest-type stars in the sample and have
a very low signal to noise ratio (which applies for the quiescent spectrum for LHS~2034 as
well) which can cause the non detection of Fe\,{\sc xiii}.  In
order to answer this question we took the spectrum of CN~Leo 
as a template and reduced it to the lower signal to noise ratio found in our DX~Cnc 
and the quiescent LHS~2034 spectra. 
The results of this exercise for DX~Cnc are shown in Fig. \ref{sn}. Clearly
the Ti\,{\sc ii} and Fe\,{\sc xiii} lines can be recognized in this example. Carrying out a 
larger number of such simulations we found
that the Fe\,{\sc xiii} need not always be recognisable, but the Ti\,{\sc ii} line should 
always be found in lower SNR spectra.
Thus the level of activity in these stars must 
be lower making the persistent presence of the Fe\,{\sc xiii} line unlikely. 

\subsection{UV Cet}

UV~Cet is a binary flare star well known from the radio to X-ray band \citep{Stepanov} and 
has been resolved in the X-ray for the first time only recently \citep{UVCetX}. 
Our UVES spectrum was taken with both components in the slit, but 
 we were unable to resolve the spectra of the
two stars (due to the seeing). Therefore the spectrum should be dominated by the brighter
A-component.  The emission lines seen in the spectrum of UV~Cet seem to have
superimposed a second component of narrow emission lines which are slightly redshifted.
We tentatively interpret this second set of emission lines as an active chromospheric 
region occupying only a part of the stars surface, since it is too narrow to be ascribed 
to the second component of the binary system.

\begin{figure}
\begin{center}

\includegraphics[width=8cm,height=5cm,clip=0]{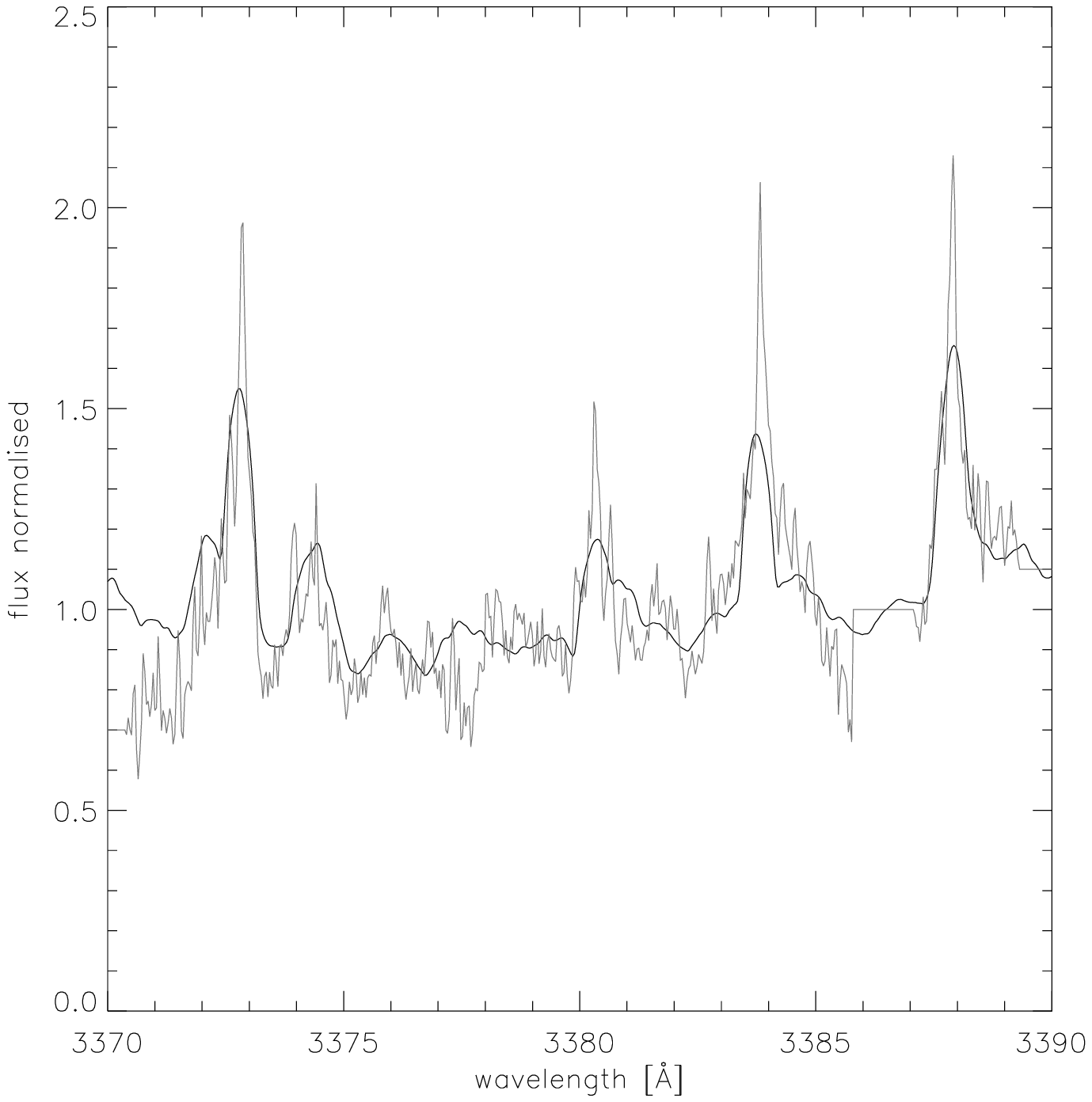}
\caption{\label{UV-Cet_comp}The spectrum of UV~Cet 3388.1 \AA\, (grey) in comparison to the
spectrum of CN~Leo (black) broadened artificially to the rotational velocity of UV~Cet.}  
\end{center}
\end{figure}

\begin{figure}
\begin{center}

\includegraphics[width=8cm,height=5cm,clip=0]{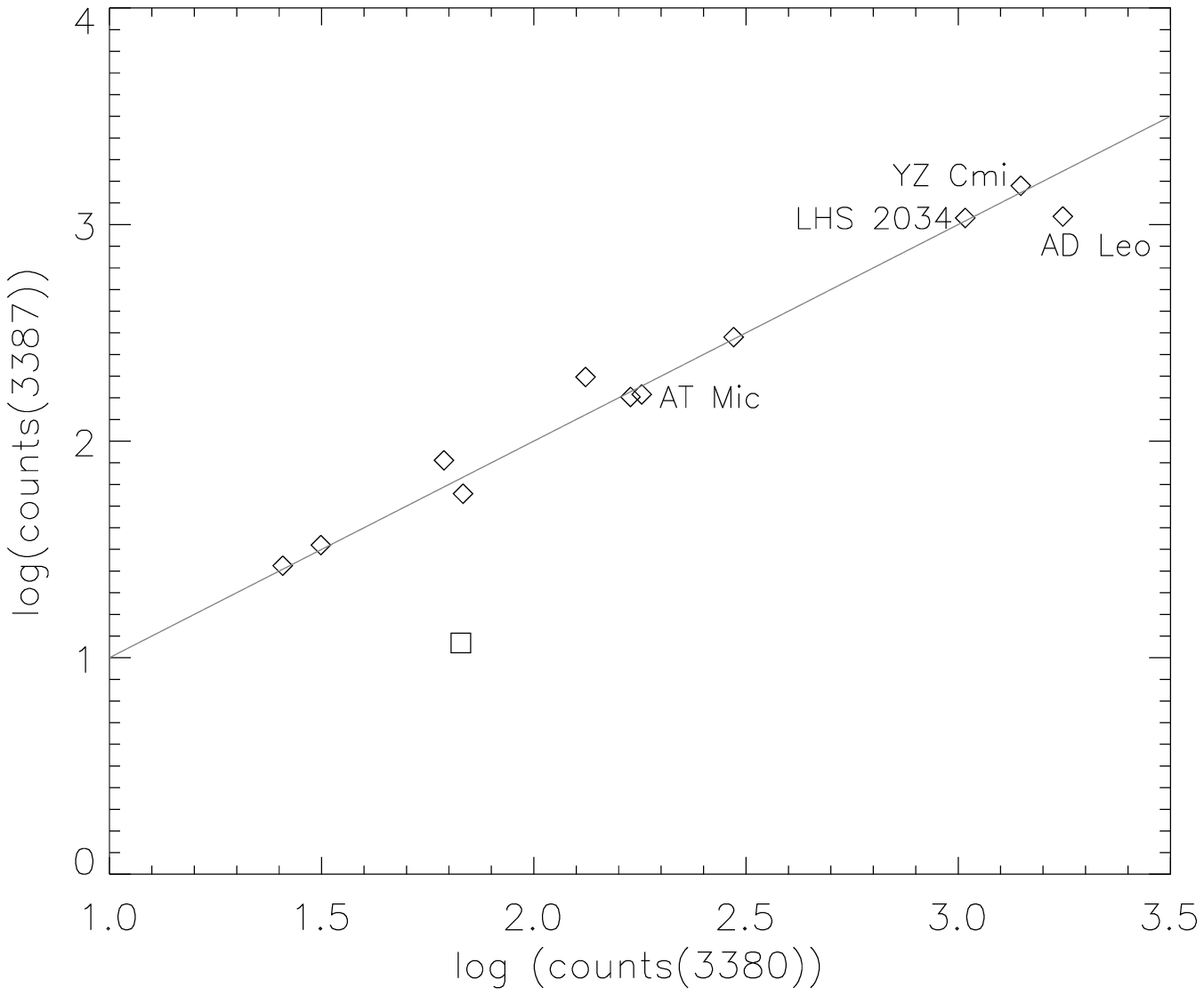}
\includegraphics[width=8cm,height=5cm,clip=0]{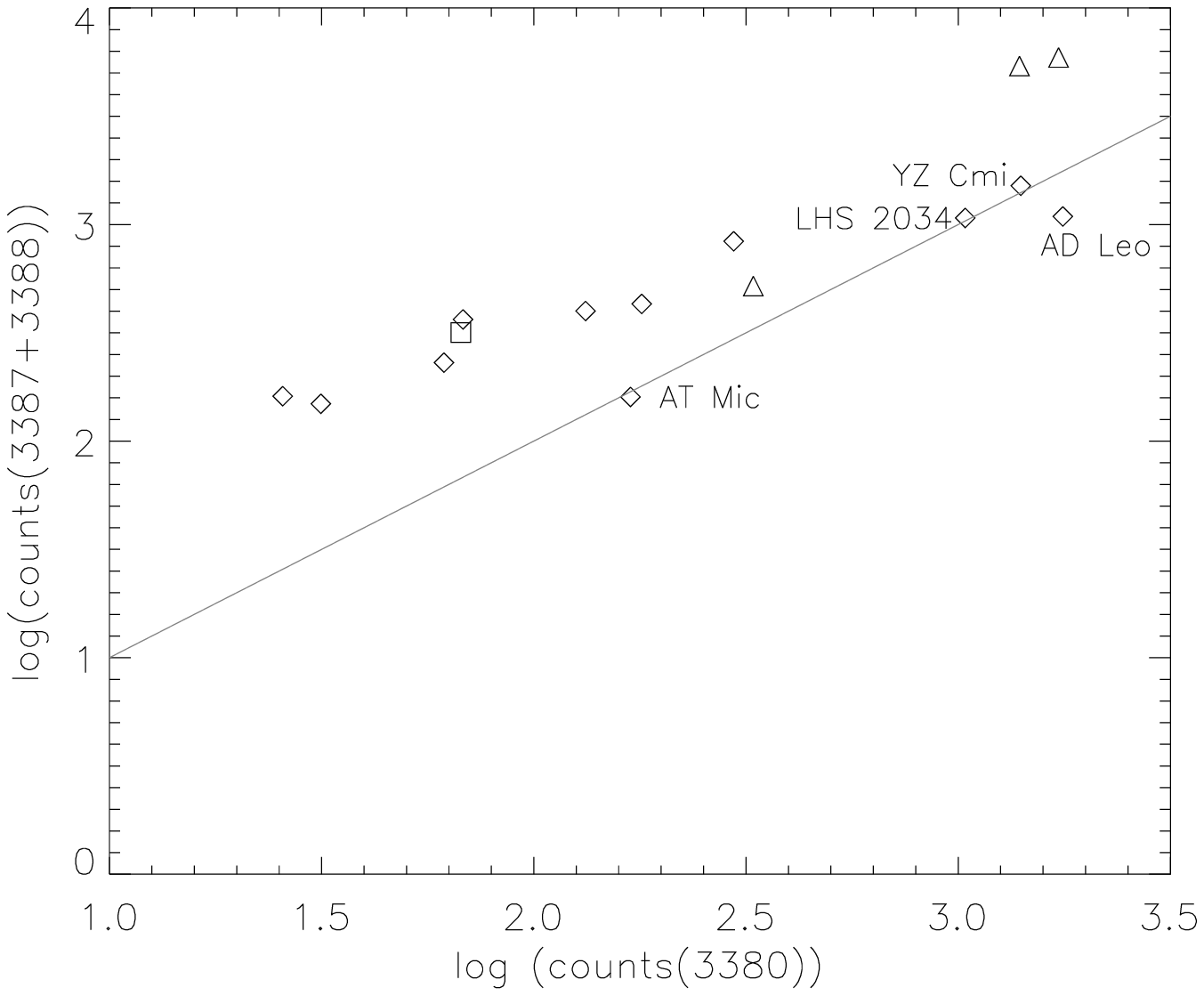}
\caption{\label{TiFeratio}Top panel: Logarithm of the 
line flux (in electrons) of the 3387 \AA\,line plotted versus the logarithm of the
line flux of the 3380 \AA\, 
line for the spectra of the CN~Leo time series, LHS~2034, LHS~2076 during the flare, YZ~CMi,
AD~Leo, AT~Mic.
The straight line marks the ratio of unity. The square marks the flux of LHS~2076
 where the line is blended with Fe\,{\sc xiii} and the fit cannot disentangle the two
lines well. Bottom panel: Logarithm of the combined flux of the 3387 \AA\, and
the Fe\,{\sc xiii} line plotted versus the 3380 \AA\, line. The triangles denote UV~Cet
with respect to the combined set of emission lines. 
Clearly AD~Leo, AT~Mic,
YZ~CMi and LHS~2034 show no Fe\,{\sc xiii} line emission, while
the UV~Cet line ratios are in the regime of stars with additional Fe\,{\sc xiii} flux (note, that
the square denoting LHS~2076 has moved above the line of unity). }
\end{center}
\end{figure}

Searching for the Fe\,{\sc xiii} line we compared the broad component of the spectrum of UV~Cet
with an artificially broadened spectrum of CN~Leo (see Fig. \ref{UV-Cet_comp}). 
The Fe\,{\sc xiii} line 
disappears in the broadened spectrum since it becomes totally blended with the
Ti\,{\sc ii} line at 3387.84 \AA. Therefore a detection of the Fe\,{\sc xiii} line in the
UV~Cet spectrum can only be obtained through indirect reasoning as follows:
First, the broadened spectrum of CN~Leo fits very well to the UV~Cet spectrum in this 
wavelength region, although - admittedly - 
in other spectral regions there is less similarity.
Second, we know from CN~Leo and atomic data that the flux in the Ti\,{\sc ii} line 
at 3387.84 \AA\, should be less than that contained in the 3372.80 \AA\, line, 
but for UV~Cet we find just the opposite.  In order to determine line fluxes,
we fitted each spectral line simultaneously with a  narrow and
a broad component (see Table \ref{UVcetprop}). If the broad component comes from the whole
surface of the star and the narrow component from an active region, the combined amplitudes 
must be considered and the combined amplitude of the 3387.84 \AA\, line is higher 
than the amplitude of the 3372.80 \AA\, line thus providing evidence of additional 
emission in the lines. The same applies for a comparison with the Ti\,{\sc ii} line at 3380.28 \AA\,
for which one expects equal flux ratios. If one plots the line ratios
for the time series of CN~Leo and other stars, one finds, that the UV~Cet lines
are the only one with a ratio that needs additional flux in the 3387.84 \AA\, line (see Fig.
\ref{TiFeratio}). 
Given the high degree of activity on UV~Cet it is suggestive
to attribute this additional emission to the Fe\,{\sc xiii} line, but given 
the complicated situation with the narrow and broad set of emission lines we
 consider the detection of Fe~XIII in UV~Cet as tentative.

\begin{table*}
\caption{\label{UVcetprop}Properties of the Ti\,{\sc ii} of UV~Cet measured with CORA using
a broad and a narrow component simultaneously. Given are always the values for the narrow
(first) and the broad component (second).}
\begin{tabular}[htbp]{ccccc}
\hline
& Ti\,{\sc ii} (3372.80 \AA) & Ti\,{\sc ii} (3380.28 \AA) & Ti\,{\sc ii} (3383.77 \AA) & Ti\,{\sc ii} (3387.84 \AA)\\
\hline
amplitude& 380 and 2416 & 329 and 1394 &662 and 4665 & 520 and 5393\\
central $\lambda$& 3372.80 & 3380.31 & 3383.82 & 3387.88\\
$\sigma$& 0.04 and 0.23 & 0.05 and 0.30 & 0.07 and 0.53 & 0.05 and 0.46\\
\hline
\end{tabular}
\end{table*}

\subsection{AD~Leo, AT~Mic, YZ~CMi, and FN~Vir} 

\begin{figure}
\begin{center}

\includegraphics[width=8cm,height=5cm,clip=0]{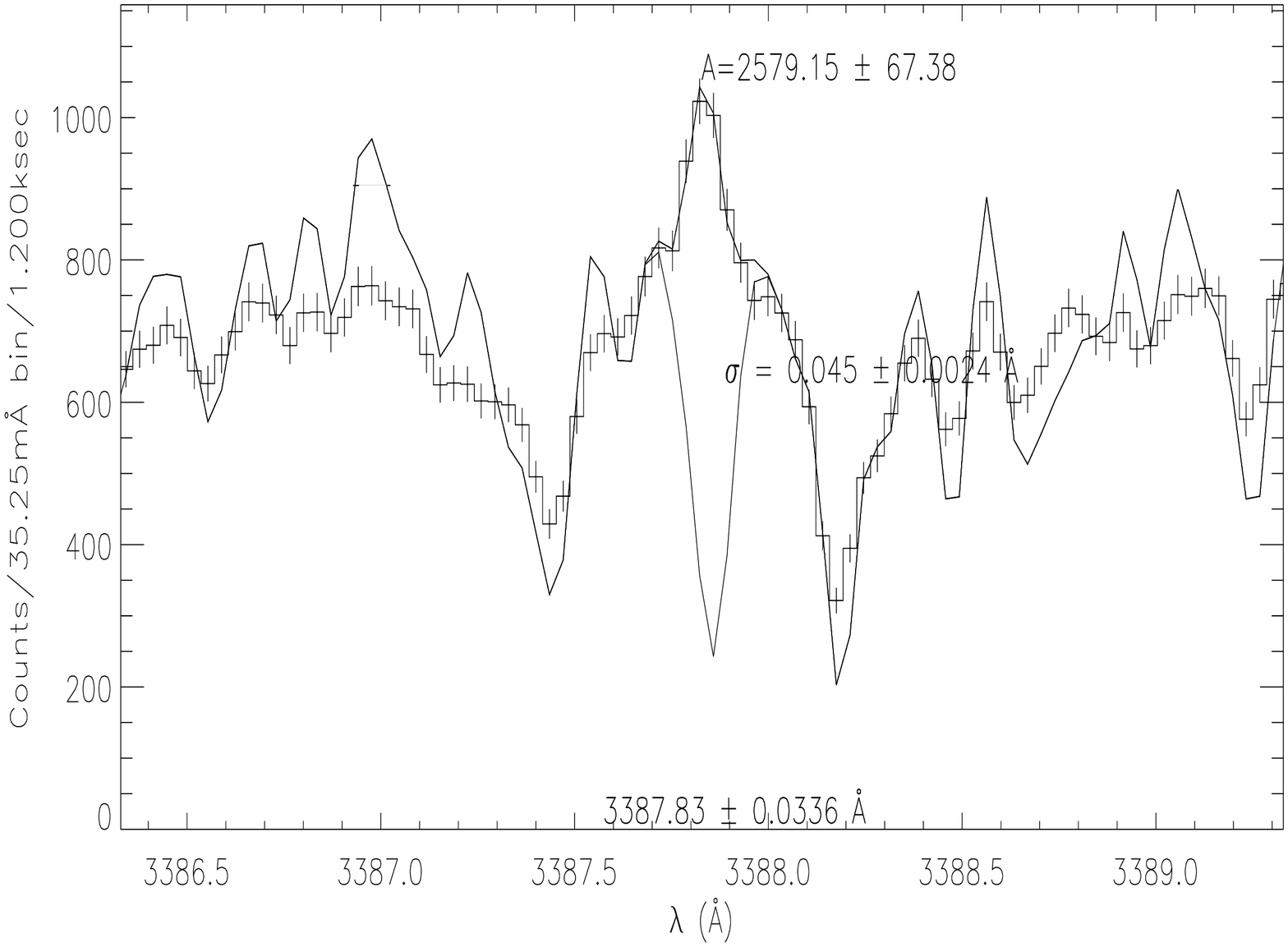}
\includegraphics[width=8cm,height=5cm,clip=0]{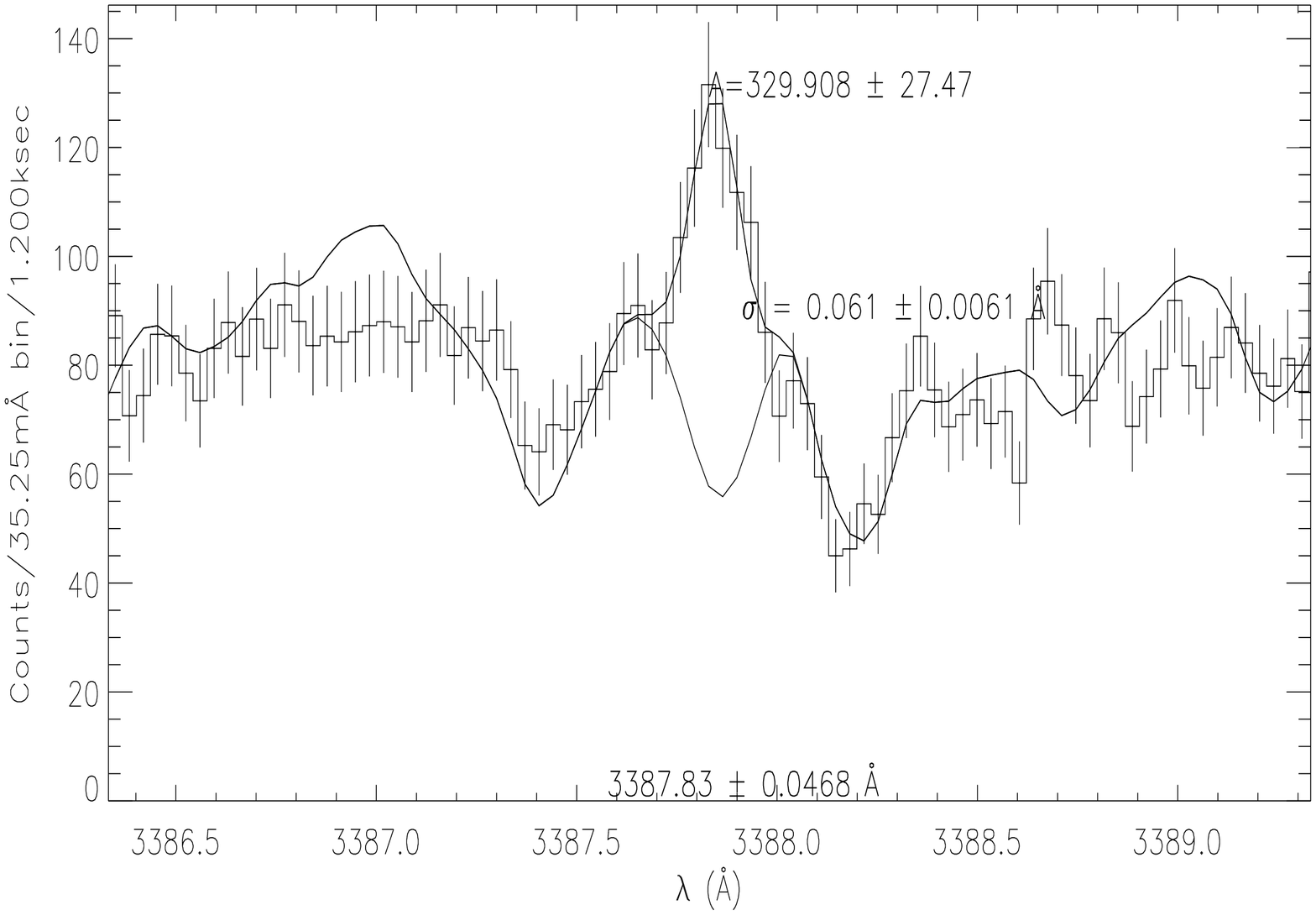}
\includegraphics[width=8cm,height=5cm,clip=0]{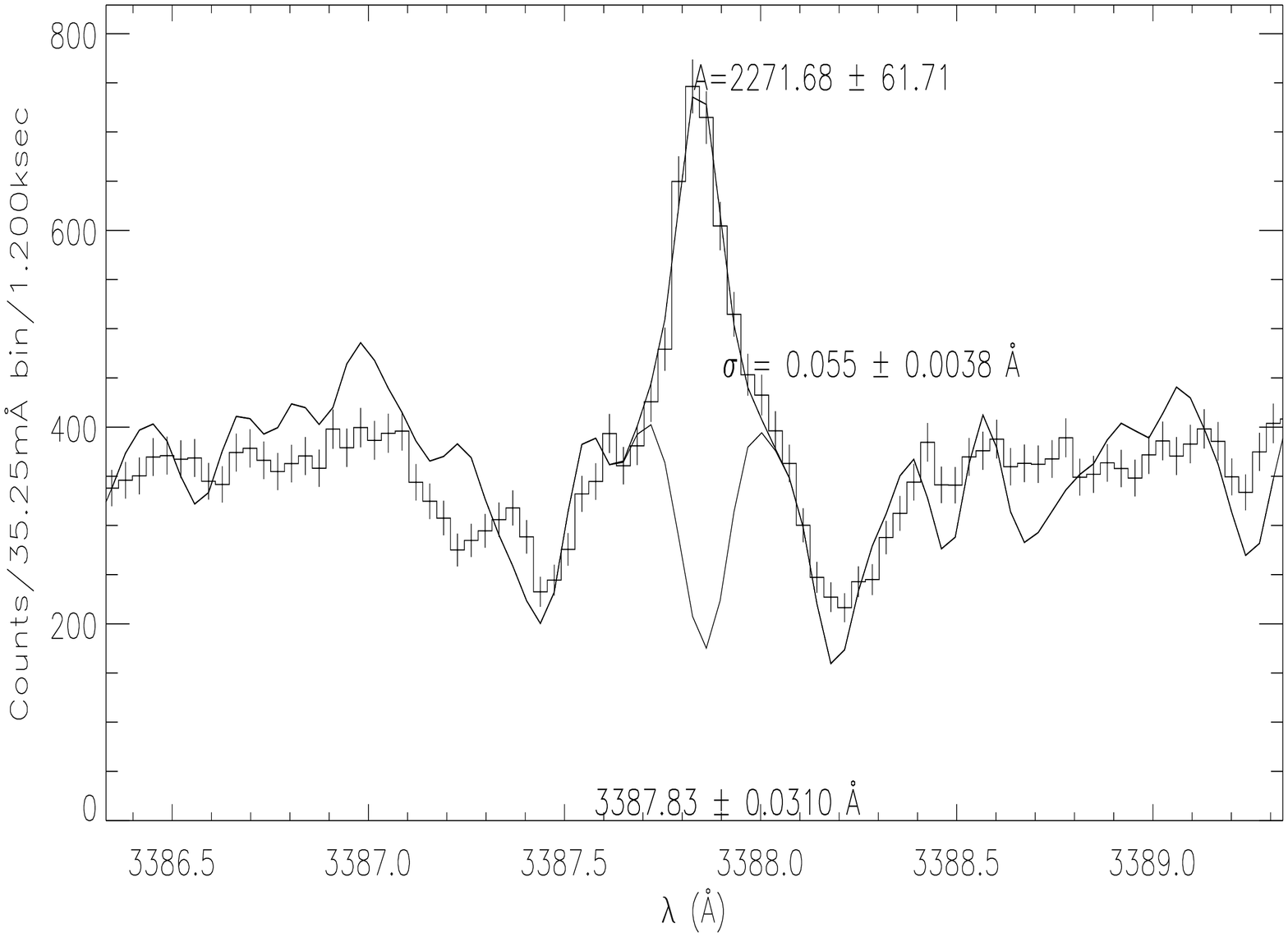}
\caption{\label{compare}The best fit of the 3387 \AA\, line for the stars AD~Leo, AT~Mic
and YZ~CMi with the rotational broadened spectra of GJ~229 as background (black line).}  
\end{center}
\end{figure}

For the other stars in our sample with spectral type M4.5 - M6 the situation is not 
as clear due to the presence of two absorption lines at 3387.45 (Ni I) and 
3388.17 (Co I) \AA\ in the spectrum. This situation applies for the three stars
AT~Mic, YZ~CMi and AD~Leo: The Ti\,{\sc ii} lines at 3372, 3380 and 3383 \AA\,  have a width of 
about 0.05 \AA\, whereas the line at 3387 \AA\, shows a width of 0.1 \AA.  This can be 
due to the Fe\,{\sc xiii} line blended by the absorption feature at 
3388.17 \AA, alternatively, leftover continuum may also be an explanation for
the broader Ti\,{\sc ii} line since the spectra of the three stars do show many absorption lines.
In order to test this hypothesis we used the spectrum of the earliest
M dwarf GJ~229 as a template for the photospheric part of the spectrum applying the
spun up and scaled spectrum  as a ``background'' for 
the emission line fit with CORA. For the three stars AT~Mic, 
YZ CMi and AD~Leo this method yields a best fit with a narrow Ti\,{\sc ii} line of 
about 0.05 \AA\, half width $\sigma$, consistent with the half width found for
the other Ti\,{\sc ii} lines in these stars. The fitted spectra are displayed
in Fig. \ref{compare}.  Therefore the apparent width of the 3387 \AA\, line seems to be 
due to photospheric continuum and there is no need to invoke
Fe\,{\sc xiii} emission in these three stars. 

For the case of FN~Vir this method cannot be used, because the template  GJ~229 deviates
significantly from the photospheric absorption lines seen in FN~Vir. However,
since the Ti\,{\sc ii} line at 3372 \AA\, is found to have the same half width $\sigma$
(0.11 $\pm$ 0.01 \AA ) as the 3387 \AA\, line
(0.12 $\pm$ 0.01 \AA ), there is again no need to invoke any Fe\,{\sc xiii} emission for this star, either. 

\subsection{Proxima Centauri}

For Proxima Centauri the Ti\,{\sc ii} lines are clearly seen in emission, but again the 
template GJ~229 deviates significantly if overlaid to the spectrum. In addition to this, the
continuum is ill defined and therefore a fit of the line very difficult although especially 
the Ti\,{\sc ii} line at 3387 \AA\, is clearly seen. 
Although the bump on the red side of the Ti\,{\sc ii} line at 3387.84 \AA\,
is somewhat suggestive of a Fe\,{\sc xiii} emission line blended with the absorption line
of Co I, this could also be due to leftover continuum. Since we were not able to
obtain meaningful fits, the situation for Proxima Centauri must remain open and
we must refrain from drawing any conclusions.


\section{Timing behavior of the Fe\,{\sc xiii} line in CN~Leo}\label{timeseries}

\begin{figure}
\begin{center}
\includegraphics[width=8cm,height=5cm,clip=]{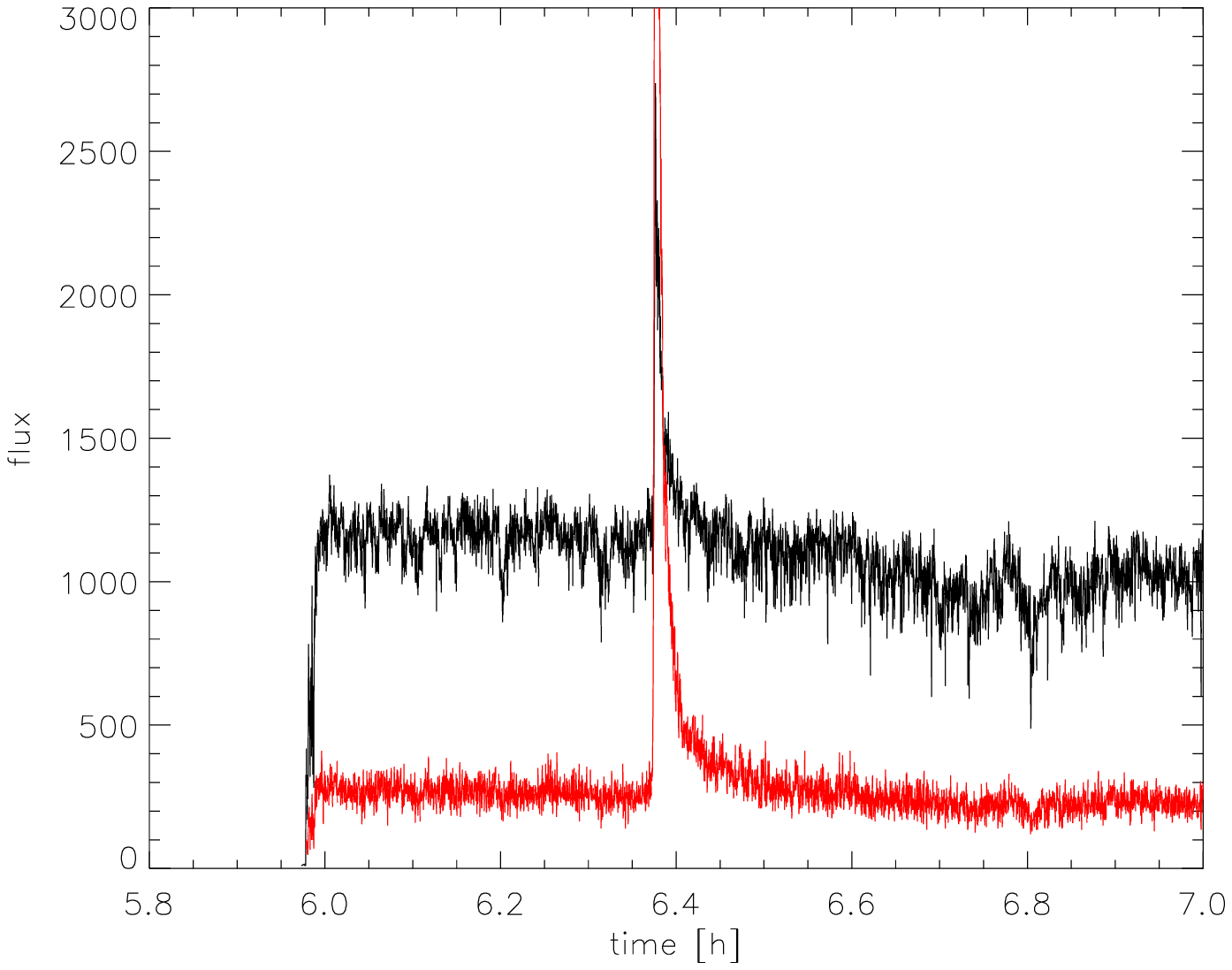}
\caption{\label{photmeter}Light curve of the flux of CN~Leo on March, 16th, during the 
second observation
block but otherwise like in Fig. \ref{photm}. Clearly a major flare can be seen 
in both the red and the blue arm of the photometer.}
\end{center}
\end{figure}

Every night during the observation run in March 2002 (with the exception of March, 14) two 
series of spectra were taken of CN~Leo, each consisting of 3 spectra with 20 minutes exposure. 
 On March, 14th only one spectral series of 
one hour duration was taken, one single spectrum taken about 2 hours
later was combined with the others to only one time series. 
We averaged the spectra of each series to improve the signal to noise ratio and 
constructed a total of 7 spectra of CN~Leo.  During the second spectral series on 
March, 16th CN~Leo a major flare occurred as can be seen in the photometer light 
curve in Fig. \ref{photmeter}; on March, 14th CN Leo was very quiet, showing very little 
of the flickering seen in the light curves on the other days.

We used the CORA line fitting program to measure the Ti\,{\sc ii} lines and 
the Fe\,{\sc xiii} line fluxes as described in section \ref{corafitting};
a listing of our fit results can be found in Table \ref{CNfluxes}.  The 
Fe\,{\sc xiii} line was detected in all spectra. However, for the spectrum 
taken on March, 16th during the flare the fit results are ambiguous. One can obtain a
good fit with two narrow lines or a narrow and a broad line, where the broad line
has nearly the same central wavelength like the narrow line (i.\,e. the
Fe\,{\sc xiii} line is blueshifted). We assume that the fit
with the lineshift of the  Fe\,{\sc xiii} line is the correct one, since there
is definitely a second line and a physical interpretation with two narrow lines one of 
which is ascribed to Fe\,{\sc xiii} with a halfwidth $\sigma$ of only 0.1 \AA\, 
is not meaningful due to temperature broadening.
Therefore a fit with two narrow
lines implies that there is no Fe\,{\sc xiii} present. This can be possible, when
the Fe\,{\sc xiii} is ionized to higher ionisation stages during the flare. This is
quite unlikely though, since we find no Fe\,{\sc xiv} emission at 5303 \AA\, in
CN~Leo (see section \ref{other}) even during the flare. Moreover the second emission
line has to be explained otherwise. We favour therefore an interpretation with a narrow Ti\,{\sc ii}
line and a broad blue shifted Fe\,{\sc xiii} line. Line shifts
are known to occur during flares, when the emitting material is raised in the
atmosphere. The velocity towards the observer in the line of sight is about 20
$\mathrm{km\, s^{-1}}$ in this case. A blueshifted Fe\,{\sc xiii} line during the
flare lines in with the blueshift of the Fe\,{\sc xiii} line 
we observed in LHS~2076 during the flare.

\begin{table*}
\caption{\label{CNfluxes}List of the line fluxes, central wavelength and half widths $\sigma$ measured with CORA. 
Due to the signal to noise ratio the computation of the error of the half width could not be done 
for every line.}
\footnotesize
\begin{tabular}[htbp]{cccccc}
spectral block &  Ti\,{\sc ii} (3372.80 \AA) & Ti\,{\sc ii} (3380.28 \AA) & Ti\,{\sc ii} (3383.77 \AA) & Ti\,{\sc ii} (3387.84 \AA) & Fe\,{\sc xiii} (3388.1 \AA)\\
\hline
2002-03-13 1st & 137.7$\pm$16.3 & 31.5$\pm$9.6 & 83.2$\pm$12.8 &33.1$\pm$10.4 &115.6$\pm$26.3\\
   &3372.77$\pm$0.05&3380.27$\pm$0.04&3383.74$\pm$0.04&3387.82$\pm$0.02&3387.88$\pm$0.17\\
            &0.06$\pm$0.01&0.03$\pm$0.01&0.04$\pm$0.01&0.02$\pm$0.01&0.20\\
\hline
2002-03-13 2nd&290.1$\pm$21.6&132.2$\pm$16.0&215.4$\pm$18.8&197.8$\pm$21.6&201.0$\pm$33.0\\
  &3372.78$\pm$0.04&3380.25$\pm$0.04&3383.73$\pm$0.03&3387.83$\pm$0.03&3388.02$\pm$0.09\\
            &0.06$\pm$0.01&0.05$\pm$0.01&0.04$\pm$0.01&0.05$\pm$0.01&0.21\\
\hline
2002-03-14&161.5$\pm$15.9&61.4$\pm$12.6&122.9$\pm$14.7&81.6$\pm$16.2&148.8$\pm$26.0\\
   &3372.78$\pm$0.03&3380.27$\pm$0.05&3383.73$\pm$0.04&3387.84$\pm$0.05&3388.09$\pm$0.26\\
           &0.04$\pm$0.01&0.05$\pm$0.01&0.04$\pm$0.01&0.05&0.30\\
\hline
2002-03-15 1st &576.4$\pm$29.1&295.7$\pm$23.4&490.6$\pm$27.1&302.2$\pm$28.1&535.1$\pm$51.2\\
   &3372.79$\pm$0.03&3380.27$\pm$0.03&3383.75$\pm$0.03&3387.83$\pm$0.03&3388.02$\pm$0.09\\
           &0.04$\pm$0.01&0.04$\pm$0.01&0.04$\pm$0.01&0.04$\pm$0.01&0.26$\pm$0.02\\
\hline
2002-03-15 2nd&437.7$\pm$24.8&179.6$\pm$18.1&315.0$\pm$21.8&164.5$\pm$20.1&265.7$\pm$39.0\\
  &3372.80$\pm$0.03&3380.27$\pm$0.03&3383.76$\pm$0.03&3387.82$\pm$0.04&3388.19$\pm$0.17\\
           &0.04$\pm$0.01&0.04$\pm$0.01&0.04$\pm$0.01&0.05$\pm$0.01&0.30\\
\hline
2002-03-16 1st&205.6$\pm$19.3&68.1$\pm$12.5&122.8$\pm$14.8&57.2$\pm$14.3&307.6$\pm$34.5\\
   &3372.80$\pm$0.05&3380.26$\pm$0.03&3383.75$\pm$0.03&3387.85$\pm$0.03&3387.98$\pm$0.17\\
           &0.06$\pm$0.01&0.03$\pm$0.01&0.04$\pm$0.01&0.03$\pm$0.01&0.24$\pm$0.02\\
\hline
2002-03-16 2nd &125.5$\pm$15.3&25.6$\pm$8.9&65.5$\pm$10.8&26.6$\pm$11.0&134.8$\pm$26.6\\
   &3372.82$\pm$0.05&3380.28$\pm$0.04&3383.76$\pm$0.03&3387.84$\pm$0.03&3387.88$\pm$0.14\\
           &0.06$\pm$0.01&0.03$\pm$0.01&0.02$\pm$0.01&0.02$\pm$0.01& 0.20\\
\hline
\end{tabular}
\normalsize
\end{table*}

\begin{figure*}[hp]
\begin{center}
\includegraphics[width=8cm,height=5cm,clip=0]{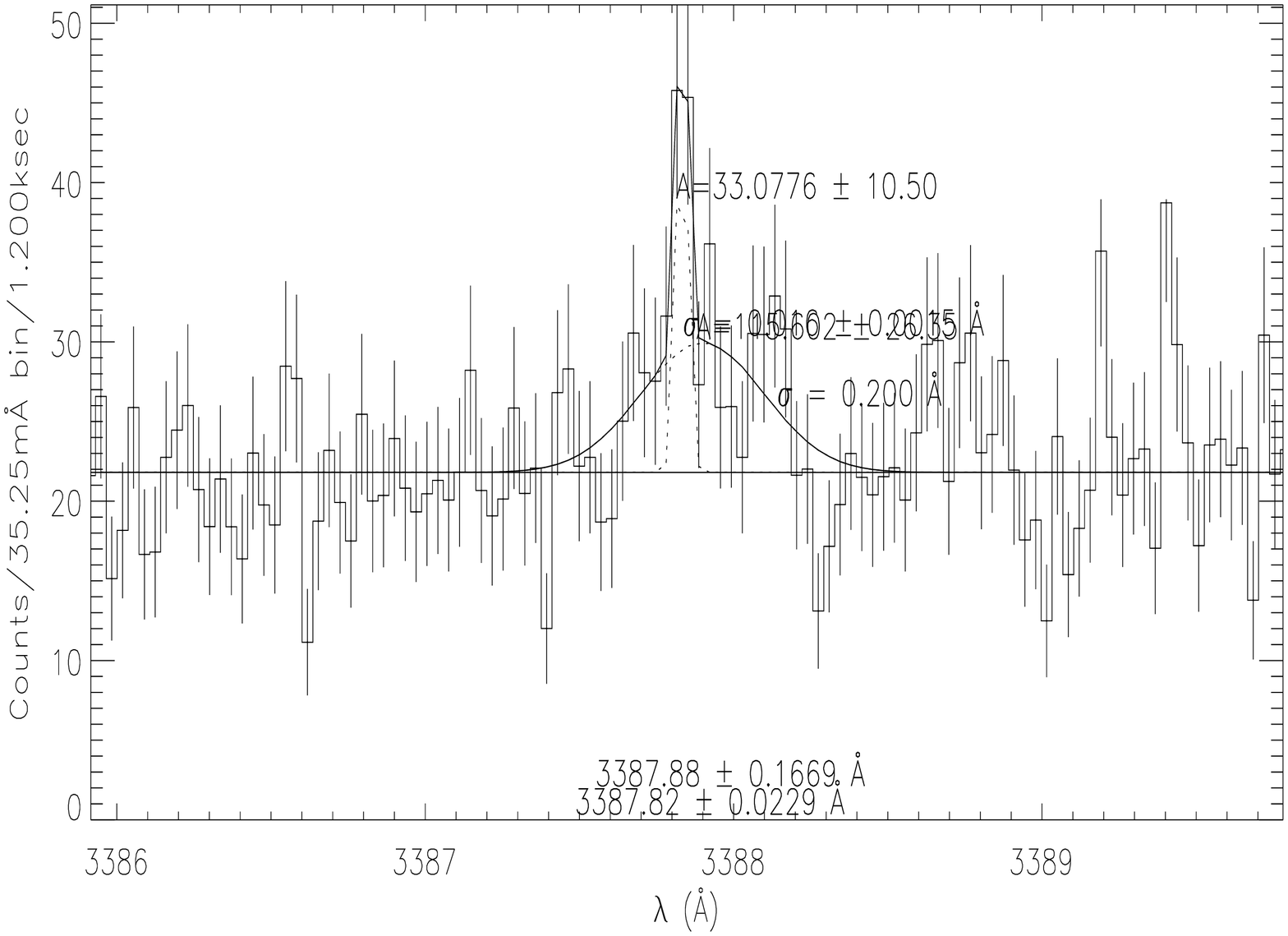}
\includegraphics[width=8cm,height=5cm,clip=0]{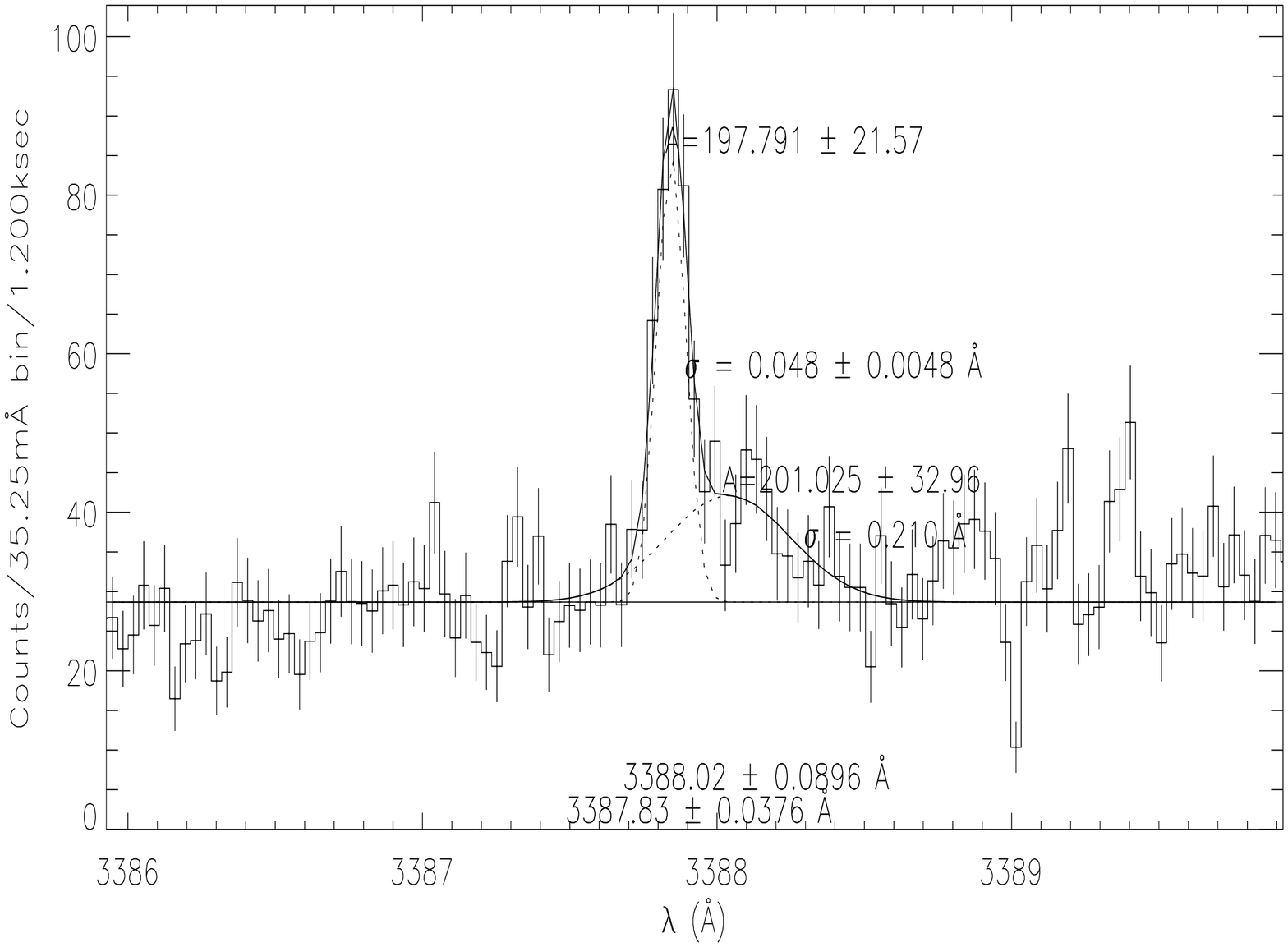}
\includegraphics[width=8cm,height=5cm,clip=0]{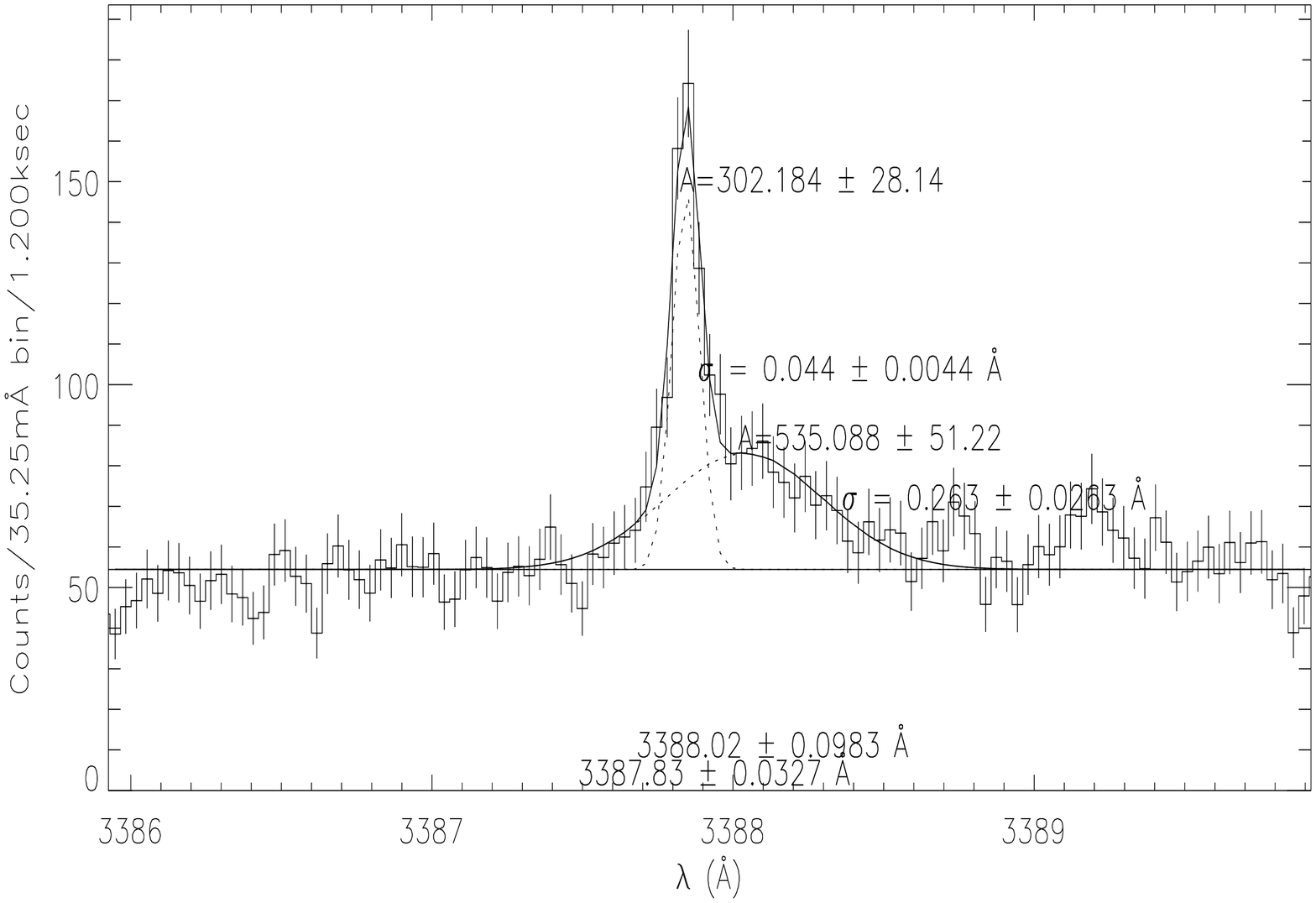}
\includegraphics[width=8cm,height=5cm,clip=0]{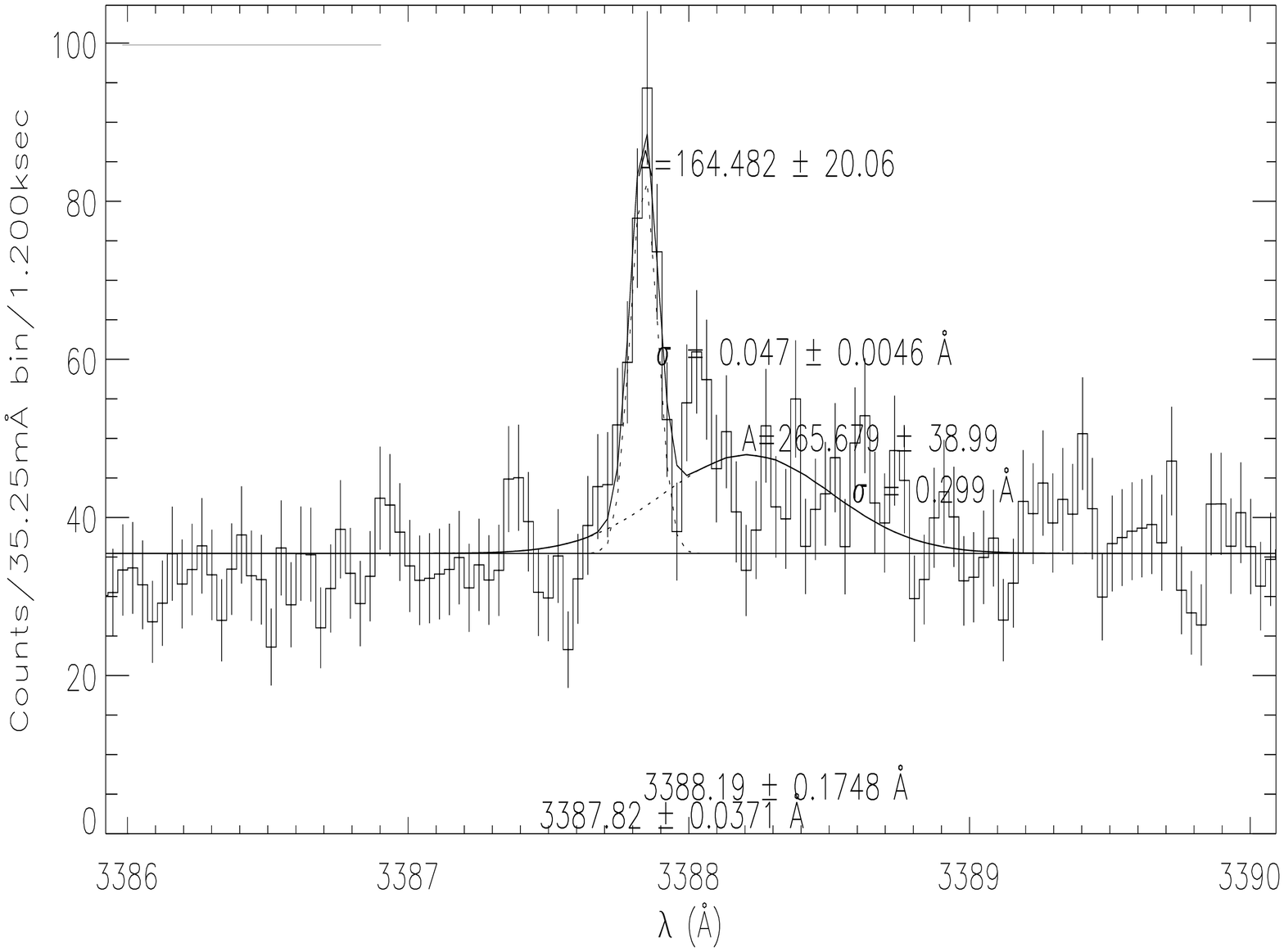}
\includegraphics[width=8cm,height=5cm,clip=0]{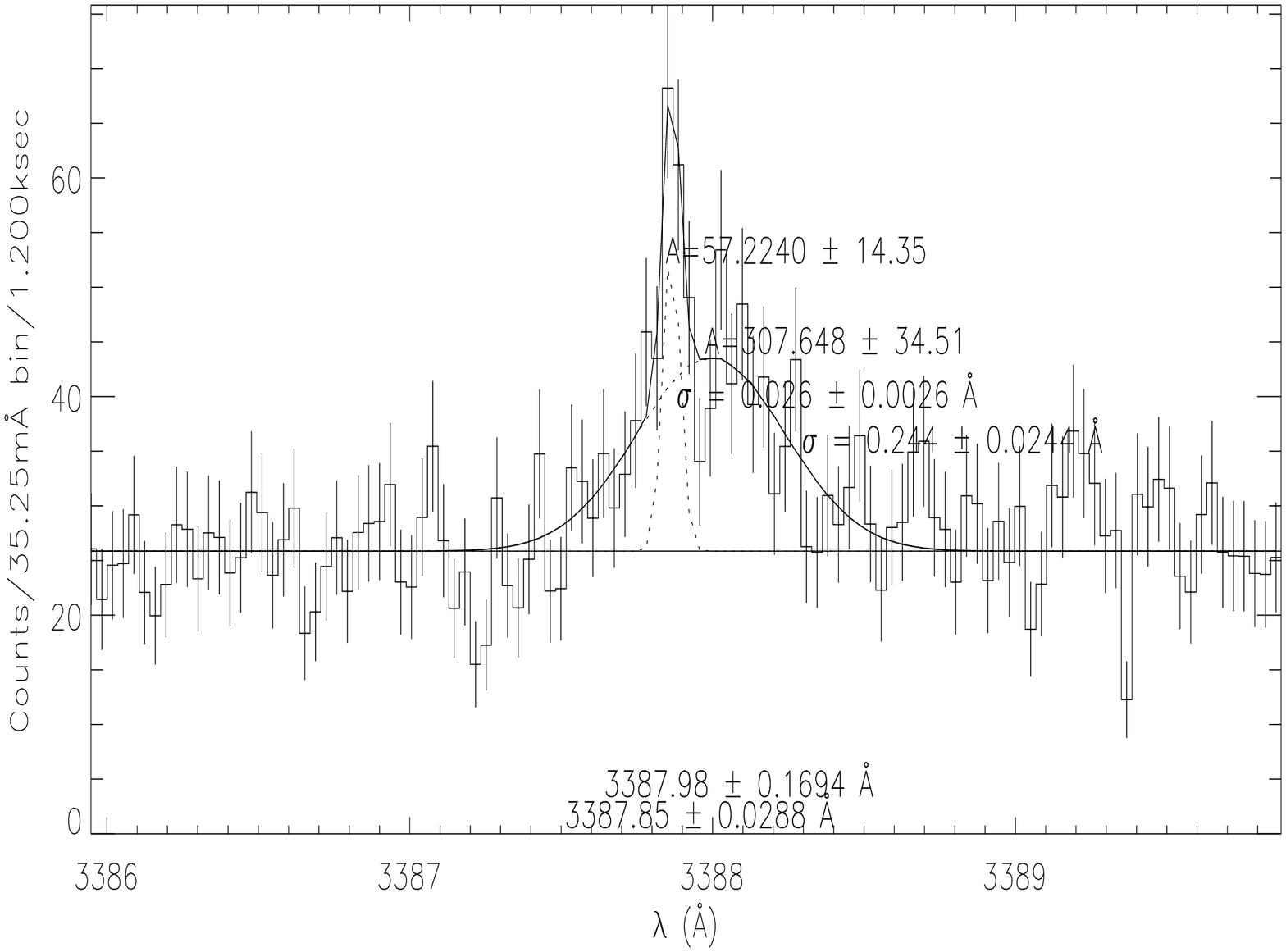}
\includegraphics[width=8cm,height=5cm,clip=0]{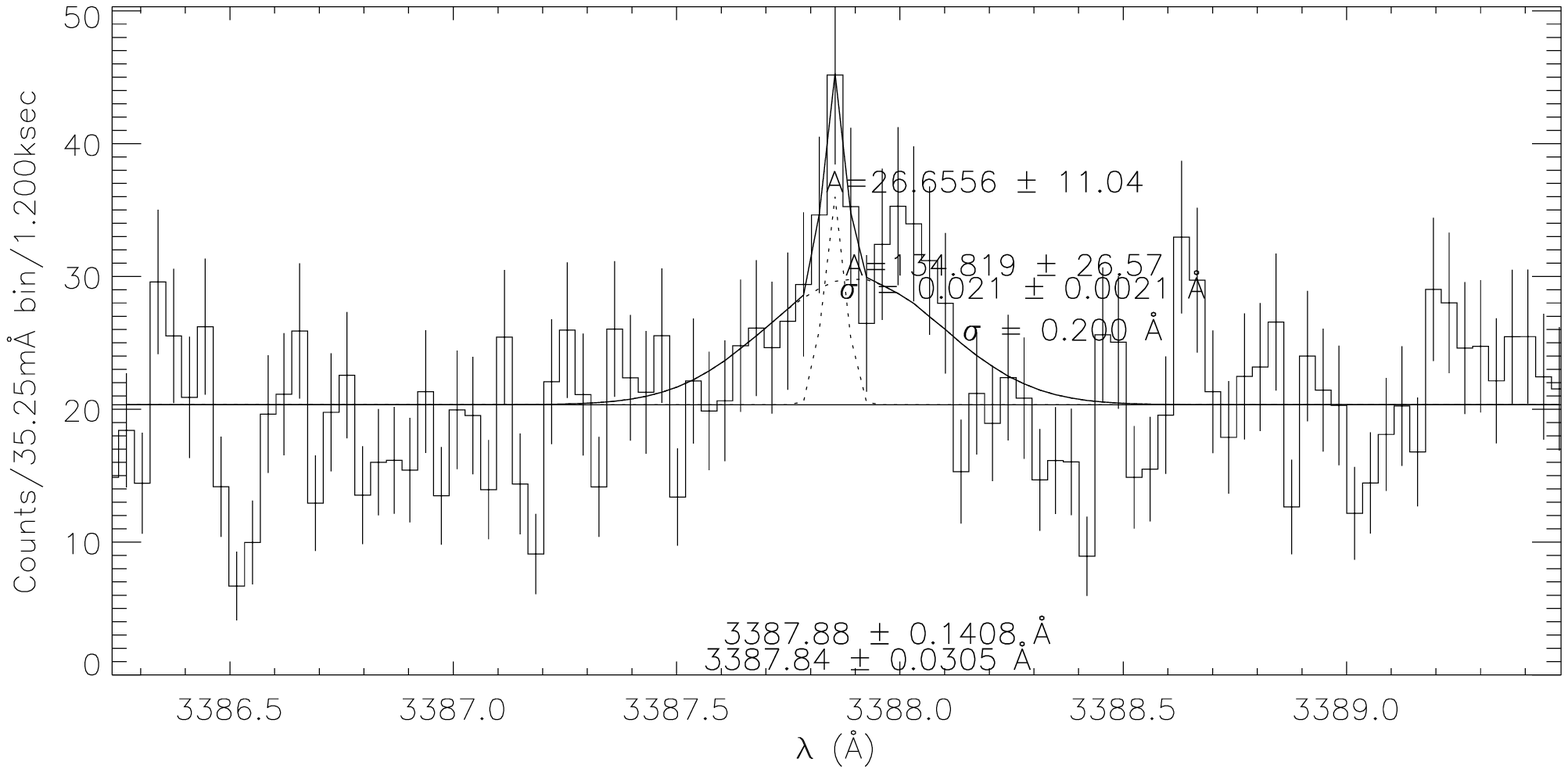}
\includegraphics[width=8cm,height=5cm,clip=0]{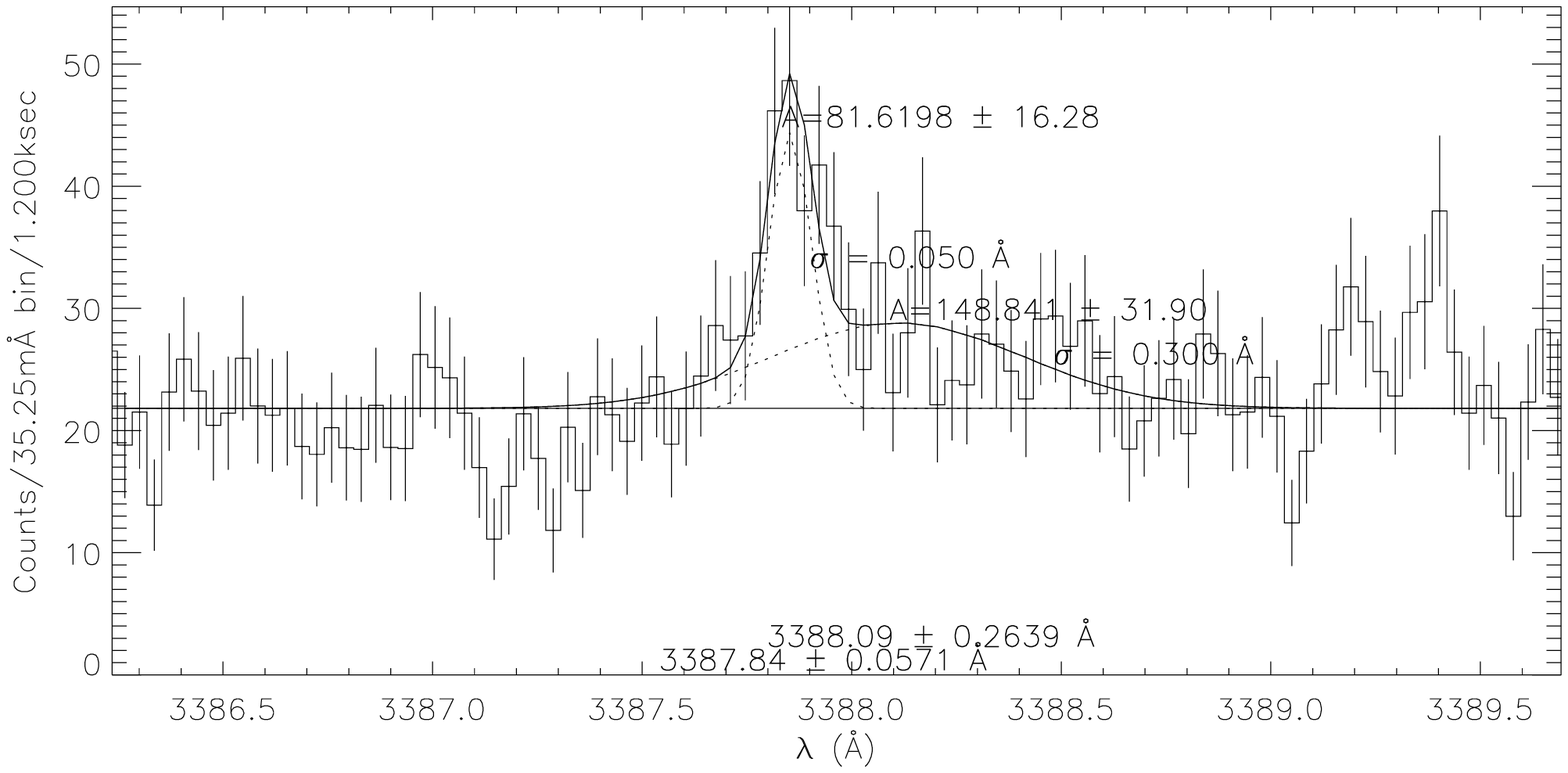}
\caption{\label{line3388}Best fit of the Ti\,{\sc ii} line together with the Fe\,{\sc xiii} line in the 
averaged spectral series. At the top the two spectra taken on March, 13th
(left: start time 2.3 UT right: start time 4.2 UT), beneath
the two spectra taken on March, 15th (left: start time 3.7 UT, right: start time 6.4 UT),
 beneath the spectra taken on March, 16th (left: start time 3.9 UT, right: start time
5.9 UT). The
single spectrum at the bottom was taken on March, 14th (start  time 3.1 UT). The horizontal
line in each spectrum indicates the background used for the line fitting.}  
\end{center}
\end{figure*}

Although the photometer data for the spectra outside the major flare are rather constant,
the corona must have undergone some variability as can be seen from Fig. \ref{line3388},
where the averaged spectra of each time series are shown.
The Fe\,{\sc xiii} line is most clearly seen in the spectrum of the first series taken 
on March, 15th, where the continuum is relatively high and the Ti\,{\sc ii} line 
is very intense, although
in the photometer data no notable flux variability is seen besides the usual flickering.
In the other spectra the line does not show that clearly, in fact the line would
be undetectable if it were only a bit weaker or if one chose a higher background 
level. Thus even for CN~Leo the Fe\,{\sc xiii} line need not to be a persistent
feature, and the high variability during quiescence is suggestive of the
microflaring as proposed for the heating of stellar coronae \citep{Kashyap}.
Also the chromospheric Ti\,{\sc ii} line shows major variability even on a timescale of hours.
For example, on March 13th the amplitude of the Ti\,{\sc ii} lines increased
by a factor of more than two over a time span of only two hours. 
Therefore also the chromosphere must have undergone rapid and large changes.


\section{X-ray and Fe\,{\sc xiii} line fluxes}

Using the standard star HD 49798 for flux calibration, we measured absolute fluxes in the 
Fe\,{\sc xiii} 3388.1 line for the stars CN~Leo, GL~Vir and for LHS 2076 (during the flare). 
Since the standard star was observed only once per night, we estimate an error in this 
absolute flux calibration up to a factor of two.  Broad band X-ray fluxes for these stars were 
obtained from broad 
band count rates and should have (systematic) errors of about 50 \%, resulting from the 
adopted count rate to energy flux conversion.  Also note that for none of our stars do we have 
simultaneous optical and X-ray observations.
The measured average fluxes for the one hour duration spectra of CN~Leo
range from $3.2\cdot 10^{-15}$ cm$^{-2}$ s$^{-1}$ in the first spectrum taken on 2002-03-13 
to $1.4 \cdot 10^{-14} \mathrm{erg\, cm^{-2} s^{-1}}$ in the first spectrum on 2002-03-15. 
For the Fe\,{\sc xiii} line flux of GL~Vir we computed $3.9 \cdot 10^{-15} 
\mathrm{erg\, cm^{-2} s^{-1}}$, and for
the flare spectrum of LHS 2076 the measured flux is $9.8 \cdot 10^{-15} 
\mathrm{erg\, cm^{-2} s^{-1}}$.  Using
distances of 2.4 pc for CN Leo 6.5 pc for GL~Vir and 5.2 pc for LHS 2076 \citep{Oppenheimer}
these fluxes result in line luminosities of
$2.0 \cdot 10^{24}$ up to $9.6 \cdot 10^{24} \mathrm{erg\, s^{-1}}$ for CN~Leo, $1.9 \cdot
10^{25} \mathrm{erg\, s^{-1}}$ for GL~Vir and 
$3.2 \cdot 10^{25}\mathrm{erg\, s^{-1}}$ for LHS 2076. Comparing these numbers with the 
non-simultaneously measured X-ray luminosities one then 
finds a ratio of broad band
X-ray to line luminosity in the range  3012 to 628 for CN~Leo, 235 for GL~Vir and of 124 for 
LHS~2076.

If one takes CN~Leo as a prototype showing Fe\,{\sc xiii} emission the 
ratios of GL~Vir and LHS~2076 are too low.  However, for LHS~2076 the Fe\,{\sc xiii} was measured
during a flare, while the X-ray luminosity was measured during quiescence and is therefore 
presumably
too low leading to an incorrect (lower) ratio. In the case of GL~Vir the line was measured
during quiescence and the low ratio adds therefore to the interpretation, that the line 
is actually the Ti\,{\sc ii} line broadened by the higher rotation rate we found for GL~Vir.

\section{Search for other forbidden coronal lines in CN~Leo}\label{other}

After the detection of the Fe\,{\sc xiii} line in CN~Leo 
the question arises if there is evidence for other forbidden coronal lines.
 A catalog of such lines can be found e.\,g. in \citet{Allen}. 
Out of this we investigated the following two lines more closely: Ca\,{\sc xii} at 3327 \AA\,
and Fe\,{\sc xiv} at 5303 \AA. 

Using the CHIANTI software package \citep{Chianti} we computed emissivity ratios for the
lines mentioned above to the Fe\,{\sc xiii} line, which are proportional to the flux
ratios of the lines. 
All computations were done assuming solar photospheric abundances since
we do not know anything specific about the coronal abundances of very low mass stars. The
ionisation balance was computed using the ionisation ratios found by \citet{Mazzotta}. The
computed flux ratios proved to be quite sensitive on the ionisation balance whose choice
can influence the computed flux ratios by more than a factor of two while the chosen
abundances have less influence.

\subsection{Analysis of the Ca\,{\sc xii} line}
\begin{figure}
\begin{center}
\includegraphics[width=8cm,height=5cm,clip=]{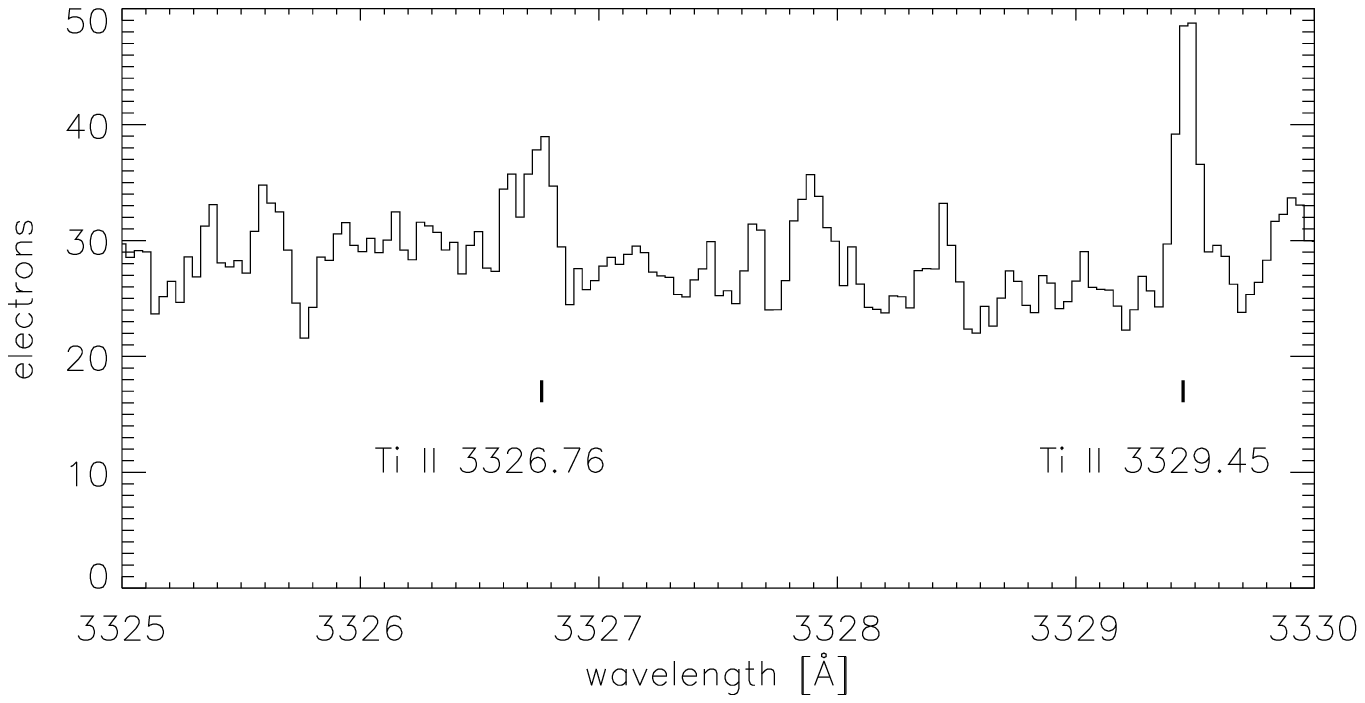}
\caption{\label{CNLeoCa}The spectrum of CN~Leo around the Ca\,{\sc xii} line. Two Ti\,{\sc ii}
lines are identified; the other faint emission features are unidentified. }
\end{center}
\end{figure}

A plot of the region around the Ca\,{\sc xii} line is shown in Fig. \ref{CNLeoCa}. Besides
the two Ti\,{\sc ii} lines the other faint emission features are unidentified, but none of them
is likely to be caused by Ca\,{\sc xii} because the line width is too narrow for a line formed
at temperatures of about log T$\sim 6$ for which one would expect a halfwidth of 0.2 \AA. 
While a line consisting of 50 counts and a halfwidth
of 0.2 \AA\, can easily be hidden
in the spectrum a line with 100 counts should be possible to detect. This leads to restrictions
for the possible temperatures and densities. Since we measured 200 counts in the Fe\,{\sc xiii} line
at 3388 \AA\, for CN~Leo averaging all our spectra,
 the highest possible ratio between the two lines that is in agreement with a 
non-detection is about 0.5, which excludes temperatures higher than log T=6.4 and lower than
log T=6.0. A more detailed discussion is given in the next subsection together with the 
restrictions drawn from Fe\,{\sc xiv}.

\subsection{Analysis of the Fe\,{\sc xiv} line}

We carefully searched for the Fe\,{\sc xiv} line at 5303 \AA. Since the line could be hidden in
the molecular lines dominating this wavelength region we compared the CN~Leo spectra to
each other and 
to the LHS~2034 averaged spectrum, assuming that this star does not display the Fe\,{\sc xiv} line. 
But no significant difference could be found. Therefore we conclude, that no 
Fe\,{\sc xiv} emission can be seen in CN~Leo.

\begin{figure}
\begin{center}
\includegraphics[width=8cm,height=5cm,clip=]{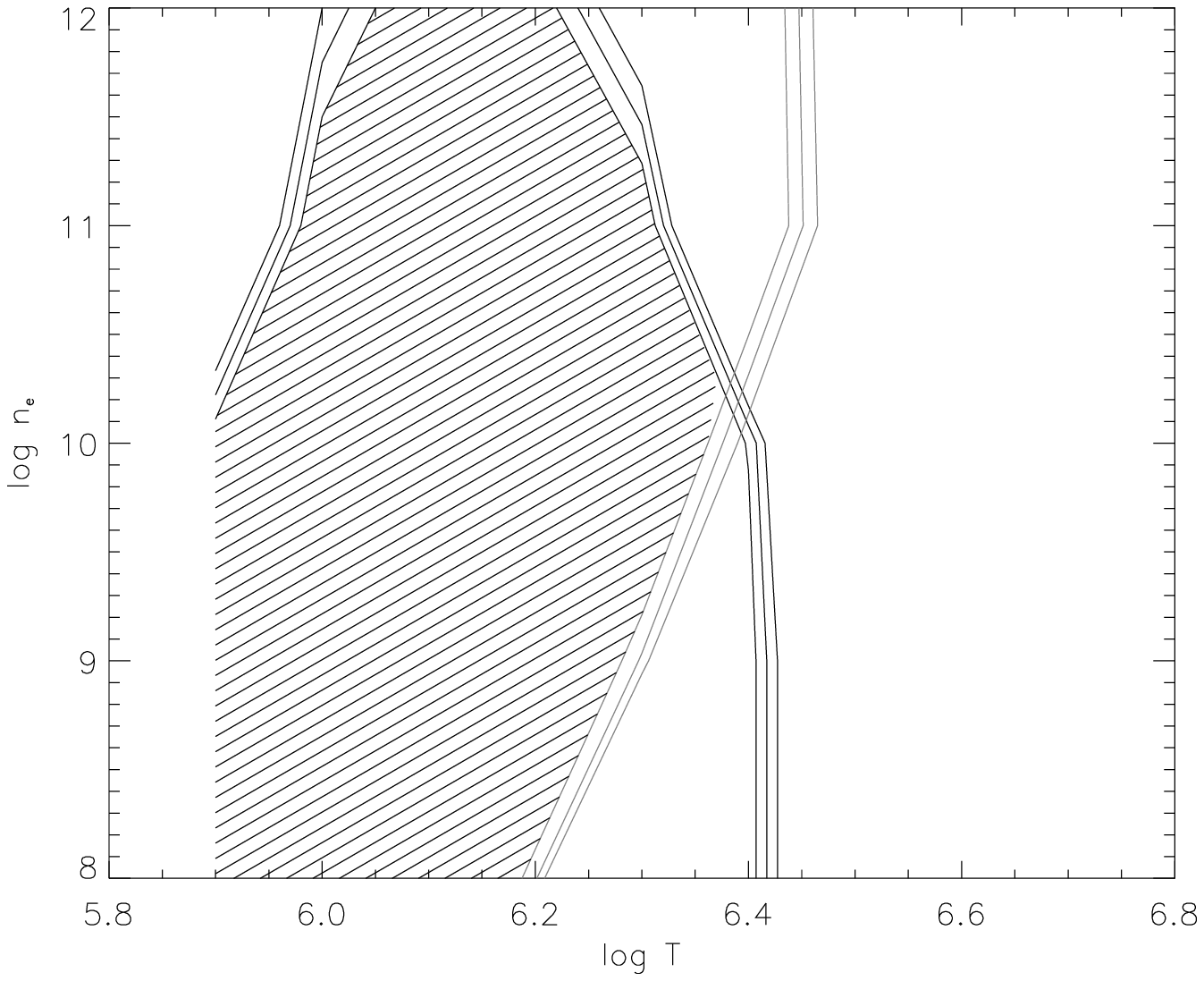}
\caption{\label{contour}Contour plot of the flux ratios between the Ca\,{\sc xii}  and
the Fe\,{\sc xiii} line (black) for a set of temperatures and electron densities and for the
ratios of the Fe\,{\sc xiv} and the Fe\,{\sc xiii} line (grey). The inner line represents a
ratio of 0.4, the middle line a ratio of 0.5 and the outer line a ratio of 0.6 for
Ca\,{\sc xii} and for Fe\,{\sc xiv} the inner line represents a ratio of 5.0, the middle
line a ratio of 5.5 and the outer line a ratio of 6.0. The shaded area denotes the
parameter set that is in agreement with the non-detection of Ca\,{\sc xii} and Fe\,{\sc xiv}. }
\end{center}
\end{figure}

Out of the absence of the Fe\,{\sc xiv} and the Ca\,{\sc xii} emission lines the possible
temperatures and densities can be estimated. To be in agreement with a non-detection for
the Ca\,{\sc xii} line the flux ratio must be lower than about 0.5 and for the Fe\,{\sc xiii} lower
than about 5.5.
In Fig.\ref{contour} a contour plot of the ratio between the
Ca\,{\sc xii} and the Fe\,{\sc xiii} and the Fe\,{\sc xiv} and the Fe\,{\sc xiii} line is given. 
From this plot one can exclude temperature
lower than log T=6.0 and higher than log T=6.2 for low densities, log T=6.3 for higher
densities and log T=6.4 for very high densities. Therefore the line formation of the Fe\,{\sc xiii}
line takes place around the peak formation temperature of log T=6.2 -- 
what would be expected from the beginning.
For the densities the plot is not very restrictive. Even densities as high as $N_{e}=10^{12} 
\mathrm{cm}^{-3}$ can be possible if the temperature is close enough to the peak formation
temperature. 

\section{Discussion and conclusions}

Our study reveals high and rapid variability of the Fe\,{\sc xiii} line on the two
stars CN~Leo and LHS 2076. While the former seems to have coronal temperatures
high enough even in its quiescent state to show the Fe\,{\sc xiii} line, the latter 
does not. Major flares were observed on both stars with opposite behavior:
In LHS 2076 we observed the Fe\,{\sc xiii} line during the flare, while in the flare 
spectrum of CN~Leo the line becomes undetectable. This can be most easily explained 
as a temperature effect: In its quiescent state the corona
of LHS 2076 is too cool to produce a sufficiently strong  
Fe\,{\sc xiii} line, but is heated to sufficient temperatures during the flare. 
On the other hand, the corona of CN~Leo is hot enough
for production the Fe\,{\sc xiii} line during quiescence, and is heated to even 
higher temperatures during flares, causing a shift of the ionization of Fe\,{\sc xiii} to 
higher ionization stages, and therefore Fe\,{\sc xiii} becomes undetectable. 

In addition to these two absolutely secure detections GL~Vir and UV~Cet may exhibit 
the Fe\,{\sc xiii} 
line, too. For Prox Cen the situation is unclear. 
In the stars AD~Leo, AT~Mic, YZ CMi and FN Vir no Fe\,{\sc xiii} line could be found. In the stars
DX~Cnc, LHS~292 and the quiescent LHS~2076 no Fe\,{\sc xiii} line could be found as well. That
may be due to the low signal to noise ratio but simulation strongly suggest that these stars are
not active enough to show the Fe\,{\sc xiii} line. Rather surprisingly
no Fe\,{\sc xiii} line could be found in the flaring LHS 2034, either.
So there are at most four detections of Fe\,{\sc xiii} in a sample of 16 or 13 stars, respectively 
if one does not count the two double stars LHS~428 and LHS~6158 and the template GJ 229A. This might
seem not very much at first glance since all are very active stars. However, one of the the 
goals of the project 
was to observe a wide range of spectral types to determine for which stars a detection of Fe\,{\sc xiii}
is feasible at all.
And if the case of LHS 2076 applies to the other
stars as well and the corona is only hot enough to show the Fe\,{\sc xiii} line during major 
flares we would only expect one more detection, since LHS 2034 showed a flare, too.
But in contrast to the two flares on CN~Leo and LHS 2076 this was a long duration
flare, whose decays lasted longer than half an hour and we would expect different
behavior of the spectral lines in the course of such a different type of flare.

Moreover the Fe\,{\sc xiii} line may have been detected for some of the stars with a
better signal to noise ratio. This can be estimated for AD~Leo where from EUVE observations
a flux at Earth of $2.69 \cdot 10^{-4} \mathrm{photons\,cm^{-2} s^{-1}}$ has been
measured for the
203.83 \AA\, line of Fe\,{\sc xiii} \citep{Micela}. Using atomic data from the CHIANTI database
\citep{Chianti}, we compute a flux ratio of the two lines of 
$\frac{F_{3388}}{F_{203}}=0.02 \dots 0.05$ assuming coronal densities in the range of $10^{8}$ 
up to $10^{10}
\mathrm{cm^{-3}}$ and temperatures in the range of 1 up to 2 $10^{6}$ K. This translates into
a line flux of $5.2 10^{-16} \cdots 1.3 10^{-15} \mathrm{erg\, cm^{-2} s^{-1}}$ in the 
Fe\,{\sc xiii} line at 
3388 \AA, which in turn would correspond to an amplitude of at most 26 counts in this 
line (for a 0.3 hr 
integration, which is the typical exposure time for all of our spectra); clearly,
such flux levels can easily be hidden in the spectrum. If the Fe\,{\sc xiii} line at 203.83 \AA\, is 
blended with other iron lines the expected amplitude would be even less. Therefore data with much better 
signal to noise are required
to settle this question. 

CN~Leo is unique in the sample when it comes to clear exhibition of the Fe\,{\sc xiii} line
in quiescent state, which we ascribe to a high level of microflaring, since the
coronal line is quite variable. This variability of the corona while the photospheric
flux measured with the photometer is quite constant calls for a parallel observation
in X-rays with Chandra or XMM-Newton, since the coronal variability should lead
to X-ray variability. 
This high level of variability in the basic coronal emission is surprising, but it finds it
counterpart in the high level of chromospheric emission
in the hydrogen emission lines of the Balmer series which are clearly seen up to
$\mathrm{H_{24}}$. This is in contrast to the Sun where even in flares the Balmer lines
are only excited up to $\mathrm{H_{16}}$ \citep{Svestka}. 

Though the variability of Fe lines for the Sun was studied extensively in X-ray wavelengths
by the satellites SOHO and TRACE the variability of the emission in Fe\,{\sc xiii} lines 
was not studied to our knowledge, yet. However for the forbidden optical green Fe\,{\sc xiv} 
line at 5303 \AA\,
there are measurements by the SOHO instrument LASCO (Large Angle Spectrometric Coronagraph) 
 \citep{Wood}. Observing quiescent coronal limb structures 
\citeauthor{Wood}
found statistically significant quasi-steady brightening on timescales of at least an hour. This is about
the same timescale on which we observed the fading of the Fe\,{\sc xiii} line in CN~Leo on March, 15th from
a very pronounced to a barely detectable line profile, which took less than three hours.


\begin{acknowledgements}
We thank Dr. Jan-Uwe Ness supplying to us his CORA line fitting program and 
providing technical support.

\end{acknowledgements}

\bibliographystyle{aa}
\bibliography{papers}

\end{document}